\documentclass[fleqn,usenatbib]{mnras}

\usepackage{newtxtext,newtxmath}
\usepackage{xcolor}
\usepackage{breqn}

\usepackage[T1]{fontenc}

\DeclareRobustCommand{\VAN}[3]{#2}
\let\VANthebibliography\thebibliography
\def\thebibliography{\DeclareRobustCommand{\VAN}[3]{##3}\VANthebibliography}

\usepackage{graphicx}	
\usepackage{amsmath}	

\title[Performance and first measurements of the MAGIC Stellar Intensity Interferometer]{Performance and first measurements of the MAGIC Stellar Intensity Interferometer}

\author[S.~Abe~et.~al.]{\parbox{\textwidth}{\Large{
S.~Abe$^{1}$,
J.~Abhir$^{2}$,
V.~A.~Acciari$^{3}$,
A.~Aguasca-Cabot$^{4}$,
I.~Agudo$^{5}$,
T.~Aniello$^{6}$,
S.~Ansoldi$^{7,43}$,
L.~A.~Antonelli$^{6}$,
A.~Arbet Engels$^{8}$,
C.~Arcaro$^{9}$
% ...
\vspace{0.4cm}\\
\textit{(Affiliations can be found after the references)}
\vspace{4.0cm}\\
}}
}

\date{Accepted XXX. Received YYY; in original form ZZZ}

\pubyear{2024}

\begin{document}
\label{firstpage}
\pagerange{\pageref{firstpage}--\pageref{lastpage}}
\maketitle

\begin{abstract}

In recent years, a new generation of optical intensity interferometers has emerged, leveraging the existing infrastructure of Imaging Atmospheric Cherenkov Telescopes (IACTs). The MAGIC telescopes host the MAGIC-SII system (Stellar Intensity Interferometer), implemented to investigate the feasibility and potential of this technique on IACTs. After the first successful measurements in 2019, the system was upgraded and now features a real-time, dead-time-free, 4-channel, GPU-based correlator. These hardware modifications allow seamless transitions between MAGIC's standard very-high-energy gamma-ray observations and optical interferometry measurements within seconds.
We establish the feasibility and potential of employing IACTs as competitive optical Intensity Interferometers with minimal hardware adjustments. The measurement of a total of 22 stellar diameters are reported, 9 corresponding to reference stars with previous comparable measurements, and 13 with no prior measurements. A prospective implementation involving telescopes from the forthcoming Cherenkov Telescope Array Observatory's northern hemisphere array, such as the first prototype of its Large-Sized Telescopes, LST-1, is technically viable. This integration would significantly enhance the sensitivity of the current system and broaden the UV-plane coverage. This advancement would enable the system to achieve competitive sensitivity with the current generation of long-baseline optical interferometers over blue wavelengths.

\end{abstract}




\section{Introduction}

The interferometry technique combines visible light (or other electromagnetic wavelengths) employing two or more telescopes to obtain a higher resolution image by applying the principle of superposition.
Radio interferometers in effect
record the electromagnetic field at disparate locations, and from
these data an aperture, which can be as
large as the Earth \citep[e.g.,][]{2022ApJ...930L..13E}, is synthesised via software. 
In the optical wavelength domain, achieving similar resolution with kilometer-scale aperture synthesis is conceivable. However, direct recording of the electromagnetic field in the same manner as in radio is currently not feasible in optical interferometry. Consequently, optical interferometry resorts to employing more indirect methods.

The best-known technique for optical interferometry goes back to
\cite{1921ApJ....53..249M} and involves bringing the light from
different telescopes together and making it interfere.  The challenge
is to maintain coherent optical paths between the telescopes.  The
early work could do so over separations of about 10\thinspace m, and
longer coherent baselines would not be realized for several decades.

In between, however, \cite{HBT1958} developed a different kind of
optical interferometry technique named intensity interferometry, which involves correlating intensity
fluctuations at different telescopes, without the need of physically combining optical beams, and not requiring coherent
baselines.\footnote{See Appendix~\ref{sec:observables} for the
interferometric observables.}  The Narrabri Stellar Intensity
Interferometer (NSII) extended the then-new Hanbury Brown and Twiss (HBT)
technique to baselines between 10 and 188\thinspace m.  Over the 1960s and
early 1970s the NSII observed 32 individual stars and 9 multiple-star
systems, measuring stellar diameters and astrometry of binary
systems \citep{HB1974}. Its limitation was not the baseline but the
signal-to-noise ratio (S/N), which prevented observing fainter sources.

Interest then returned to the Michelson-Pease type of optical
interferometers, which eventually also achieved baselines of hundreds
of metres, without the S/N limitations of HBT
interferometry.  CHARA and VLTI are the best known of these, and have
had many successes, which we will briefly discuss below. Extensions to
km-scale baselines are under development \citep[see
  e.g.,][]{2020A&A...639A..53B} with even more ambitious proposals
involving going to space or the Moon \citep{2021RSPTA.37990570L}, but
none of these projects are imminent.

For optical interferometry at km-scale baselines a nearer-term
prospect is intensity interferometry with new generation
instrumentation. In particular, implementation as a second observing
mode in Imaging Atmospheric Cherenkov Telescopes (IACTs) has been
advocated for some time \citep{2006ApJ...649..399L,Dravins2013}, and
the last few years have seen a revival of intensity interferometry.
Early results from MAGIC have been reported in \cite{magic_2019} and
\cite{spie2022}.  VERITAS, following the initial implementation of
intensity interferometry \citep{VERITAS}, has announced an ongoing survey of
stellar angular diameters \citep{VERITAS3}.  H.E.S.S. has developed
intensity interferometer \citep{HESS1}.  In addition to IACTs, results from intensity
interferometry through standard optical telescopes have also been
reported \cite[e.g.,][]{SCSI,Nice}.

The recent scientific results from HBT interferometry have been mainly
stellar radii and their implications.  These are modest compared to
what Michelson-Pease interferometry has achieved over recent years \citep{Eisenhauer_review}. Here we list some examples of recent scientific achievements led by the technique, with a special focus on those topics in which intensity interferometers may contribute: 
1) Radial oscillations of Cepheids have been spatially resolved by
  CHARA \citep[e.g.,][]{2016A&A...593A..45N} and intensity
  interfometry has the potential to resolve these and more complicated
  asteroseismic modes.
2) For fast-rotating stars, CHARA has resolved
  \citep[e.g.,][]{2011ApJ...732...68C} rotational flattening
  accompanied by gravity darkening of the equator compared to the
  poles. As shown by \cite{2015MNRAS.453.1999N}, the simultaneously available UV coverage provided by intensity interferometry telescope arrays is ideal for this science case.
3) The surface of Betelgeuse has been resolved with the VLTI
  \citep{2021sf2a.conf...13M}. Image reconstruction algorithms
  have also been adapted to intensity interferometry
  \citep{Dravins2012}, and a combination of both techniques would lead to a wider resolution coverage.
4) Especially well known is the astrometry of relativistically
  moving stars in the Galactic-centre region using VLTI
  \citep[e.g.,][]{2022A&A...657A..82G}.  
5) More ambitiously, gravitational-wave-emitting binaries have also been
  discussed as possible targets for intensity interferometry
  \citep{2020MNRAS.498.4577B}.

In order to demonstrate the viability of these scientific objectives and justify farther pursue of this technique, the performance of the current
instrumentation needs to be studied in more detail, as well as exploring the systematics that may affect instruments far exceeding the sensitivity of the NSII.

\section{Hardware setup}

MAGIC is a system of two IACTs located at the Roque de los Muchachos Observatory on the island of La Palma in Spain \citep{Aleksic2016}. Equipped with 17~m diameter parabolic reflectors and fast photomultiplier (PMT) Cherenkov-imaging cameras, the telescopes record images of extensive air showers in stereoscopic mode, enabling the observation of very-high-energy (VHE) gamma-ray sources at energies of few tens of GeV up to tens of TeV \citep{performance_upgrade}. 

In April 2019, to prove that MAGIC was technically ready to perform intensity interferometry observations, a test was performed using MAGIC telescopes and an oscilloscope as a readout \citep{magic_2019}. Temporal correlation was detected for three different stars of known angular diameter. The sensitivity and the degree of correlation were consistent with the stellar diameters and the expected instrumental parameters. 
However, the acquisition was constrained by a low duty cycle, and the mechanical installation of necessary optical filters hindered a swift transition between VHE and interferometry observations.

\begin{figure*}
   \begin{center}
   \begin{tabular}{c} 
   \includegraphics[width=0.8\textwidth, trim={1cm 0 0 0},clip]{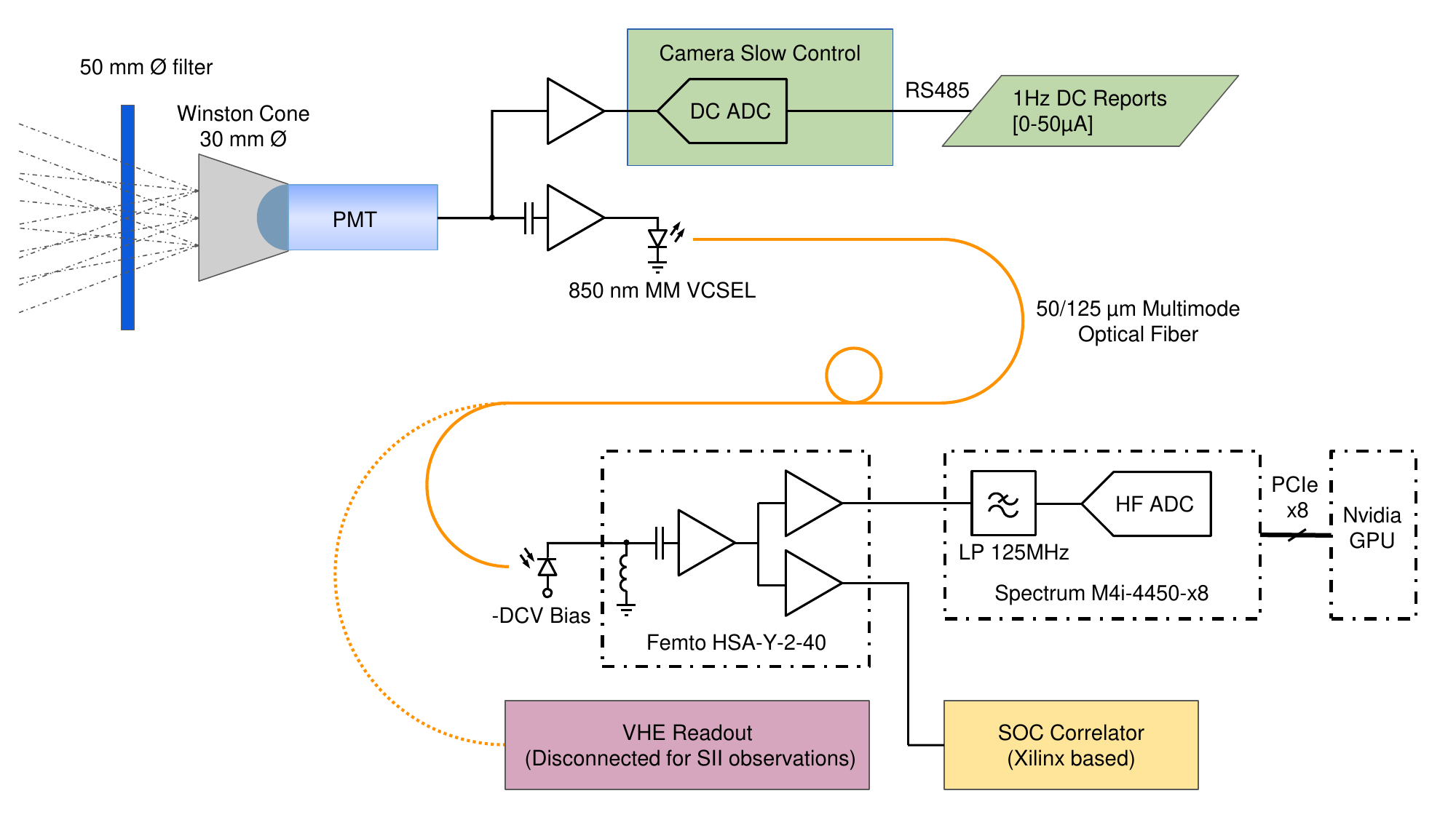}
   \end{tabular}
   \end{center}
   \caption[transmission-curve] 
   { \label{fig:sii_setup} 
Diagram describing the different components of the MAGIC-SII setup. The main differences with respect to standard VHE observations are the incorporation of narrow-band optical filters positioned in front of MAGIC photodetectors and the continuous digitization and correlation of their recorded signals with the fastest bandwidth possible, performed by a set of digitizers and a GPU-based correlator.}
   \end{figure*}

In the following we describe the technical modifications that have been implemented in MAGIC to enable intensity interferometry observations, summarised with a diagram in Fig. \ref{fig:sii_setup}. Some of these modifications have already been discussed in detail in \cite{spie2020, icrc2021, spie2022}. A key consideration was to not affect regular VHE observations, allowing a smooth and effortless transition between  ``VHE observation mode'' to ``interferometry observation mode'' and back in less than one minute. Since the implementation of these modifications in 2021, a total of 192 observing hours have been performed between January and December of 2022, predominantly during bright Moon-light periods.

\subsection{Photon detectors and signal transmitters}
\label{sec:photon_transmission}

MAGIC telescopes are equipped with 1039-pixel PMT cameras at their primary focus.
The PMTs are 25.4~mm in diameter and have 6 dynodes. 
A hexagonal shape Winston Cone is mounted on top on each PMT. The distance between PMT centers is 30~mm and corresponds to a 0.1$^{\circ}$ FOV.

The camera and data acquisition of the MAGIC telescopes were developed for efficient detection, recording, and offline reconstruction of temporally brief VHE gamma-ray events \citep{MII_daq}. The detection of low-energy events on a ns time scale against the constant night sky background (NSB) requires a wideband, low dispersion signal path from the camera focal plane to the data acquisition trigger and readout. The camera uses fast PMTs with high quantum efficiency (QE) in the wavelength range of incident Cherenkov light to respond to events while being less sensitive at NSB wavelengths. A dedicated hardware trigger allows the extended 2D image data to selectively record only when  signal is present \citep{trigger}. A high dynamic range of $10^3$ supports energy reconstruction over a wide range of incident energies. Slow control monitoring of several operating parameters including the PMT anode direct currents (DC) is done continuously during observations.

As a summary, the specifications of the camera and data acquisition for VHE observations are:
\begin{itemize}
    \item High detector QE at Cherenkov wavelengths, suppressing NSB.
    \item Wideband, low dispersion signal path from the camera to the separate readout.
    \item Low noise in the signal path to facilitate trigger response to few-photon signals.
    \item Isochronicity at the ns level across all channels.
    \item Selective recording of pixelized 2D Cherenkov images with a dynamic range of $\sim 10^3$.
    \item Continuous PMT DC monitoring at one second intervals (for monitoring purposes).
\end{itemize}

In contrast, SII observations require a continuous recording of optical signals from a point-like source over observation times ranging from minutes to hours, with a different set of desirable characteristics:

\begin{itemize}
    \item High detector QE in the wavelength band of interest.
    \item Wideband, low dispersion signal path from the camera to the separate readout.
    \item Low noise in the signal path compared to the total starlight signal.
    \item Very low correlated noise in the signal path.
    \item Continuous recording of a few pixels per telescope with moderate dynamic range and fixed temporal delay between channels (low jitter).
    \item Continuous PMT DC monitoring at one second or less intervals.
\end{itemize}

The dynamic range necessary to sufficiently resolve the steady-state input signal is dictated by the expected photon flux divided by the duration of each photon signal. This dynamic range is typically more than an order of magnitude less than that implemented for VHE observations. To aggregate measurements across varying flux levels, an essential prerequisite is a quantification of the input photon flux. In this context, we rely on the slow control DC reports, which encompass system logs storing information on PMT average photon currents and the applied high voltage (HV) over time.

Aggregation of measurements at different flux levels requires a measure of the input photon flux, for which we use the slow control DC reports (available system logs storing PMT average photon currents and applied high voltage (HV) as a function of time).

Instead of having additional photo-detectors mounted on top of their cameras (as done by VERITAS or H.E.S.S.), MAGIC-SII observations are performed with the same pixels employed for gamma-ray observations. Only a few selected camera pixels and their optical analog signal transmission are used in conjunction with a separate signal receiver and digitizer optimized for SII in place of the VHE readout. The used camera pixels contain a PMT and wideband AC coupled amplifier which drives vertical cavity surface emitting lasers (VCSELs) operating at 850~nm for analog optical signal transmission. The characteristics of the camera pixels and their optical transmission are detailed in \cite{performance_paper}.

The SII receiver consists of a multimode fiber coupled, battery reverse-biased photodiode with responsivity suitable for 850~nm and a bandwidth of 4 GHz. The detector is placed in an radio frequency (RF) enclosure coupled to the input of a Femto HSA-Y-2-40 wideband amplifier which supplies an AC coupled output signal level of $\sim$ 15 mV amplitude per single phe (dependent upon PMT gain settings) to a 50 $\Omega$ semi-rigid coaxial connection to the digitizer. The Femto amplifier has two separate output buffers, facilitating simultaneous operation of two different digitizers or concurrent signal monitoring.

The noise contributions originating from the electronics can be classified in two different categories depending on their contribution to the resulting correlation and its uncertainties:

\begin{itemize}
    \item \textit{Uncorrelated noise:} The VCSEL used in analog optical transmission contributes as relative intensity noise (RIN). Since the signal variation is small compared to the bias current of the VCSEL, it can be considered to be fixed and independent of the input signal level. The VCSEL RIN is also subject to random variations in amplitude and spectrum. The pixel preamplifier and Femto amplifier used in the receiver both contribute with additional wideband uncorrelated noise to the output signal. This noise contribution is of fixed amplitude and is smaller than the VCSEL contribution. Their combined contribution can be disregarded in the analysis provided the total optical signal power per measurement interval is greater than the total uncorrelated noise power from the electronics, otherwise it needs to be dealt with as a correction term when calculating the uncertainty. Other sources of uncorrelated noise may be pickup local to a particular telescope camera or an individual digitizer channel.

    \item \textit{Correlated noise:} Channel to channel crosstalk is an example of delay-dependent correlated noise, which is heavily correlated near delay zero between channels and uncorrelated otherwise. In our setup we avoid crosstalk contributions by introducing a fixed optical delay on the order of 500 ns in adjacent digitization channels to locate the signal outside the region of disturbance. The SII optical receivers use RF shielding, battery photodiode bias, separate amplifier power adapters, and coaxial connecting cables with effective shielding to suppress crosstalk between receivers. External signals due to a source common to multiple channels such as Radio Frequency Interference (RFI) pickup in a common camera or adjacent computing equipment, may be transient or continuous in nature and independent of the delay between channels. Their contribution may render the analysis ambiguous whenever they become comparable to the analysis uncertainty or signal amplitude.  
\end{itemize}

The pixel preamplifier, optical signal transmission, and optical receiver have a combined transfer function magnitude bandwidth measured to be $>$ 400 MHz with low dispersion using an R+S ZVA-8 vector network analyzer. Due to this, the signal channel bandwidth will be mostly determined by the PMT pulse time response ($\sim$ 2 ns FWHM) and the antialiasing filter of the digitizer (125 MHz). The aggregate bandwidth of this setup is fairly consistent with the one used by \cite{magic_2019}, approximately 110 MHz. Improving the bandwidth of the system will be discussed in section \ref{sec:future}.

\begin{figure}
\begin{center}
\begin{tabular}{c} 
\includegraphics[height=5cm, trim={1cm 2cm 1cm 2cm},clip]{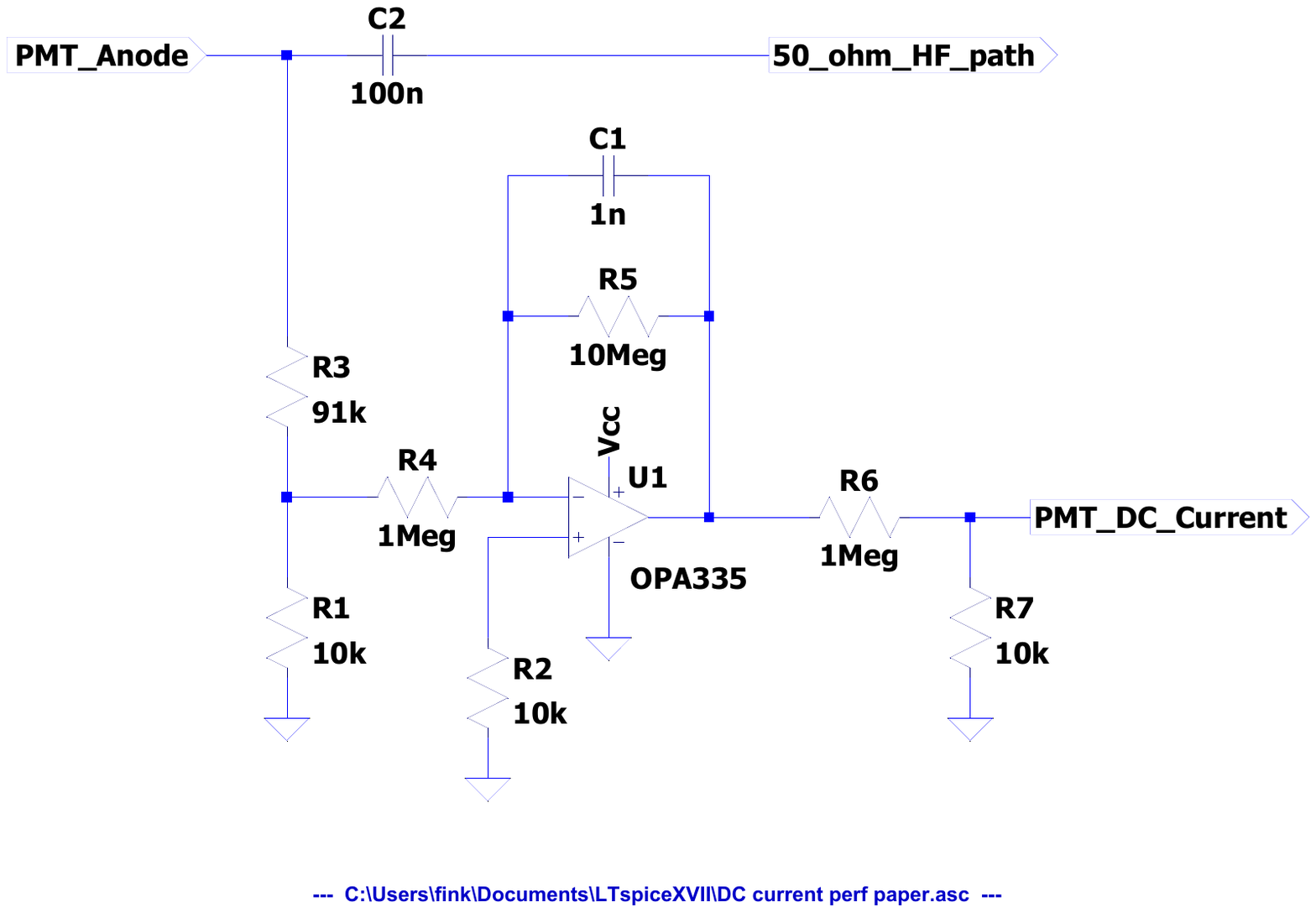}
\end{tabular}
\end{center}
\caption[DC-branch] 
{ \label{fig:DC_branch} 
Electronic circuit design of the DC monitoring path.}
\end{figure}

The DC measurement branch is used as implemented for monitoring purposes in the MAGIC cameras. Since its properties are critical to an accurate determination of the relative incident photon flux, a detailed description is provided. As shown in Fig. \ref{fig:DC_branch}, the DC monitoring path consists of a resistive divider connected directly between the PMT anode and signal ground potential. The divider decouples the DC monitoring path from excessive PMT output levels and isolates the RF signal path from the monitoring circuit. The PMT current is sensed as a (negative) voltage at the output of the divider, which is amplified by a low offset, low temperature and time drift, low input bias current operational amplifier (OPA335, Burr -Brown from Texas Instruments) in inverting configuration.

The output of the sensing circuit is connected to a 12-bit ADC with internal reference and temperature compensation, MAX1231 from Analog Devices (formerly MAXIM IC), and read out via serial interface at 1 second intervals. The reference ground for all monitored values is separate from the pixel power return line to preclude voltage drops along the connection lines.

\subsection{Optical filters mounts}
\label{sec:optomechanics}

\begin{figure}
\begin{center}
\begin{tabular}{c} 
\includegraphics[width=0.95\linewidth]{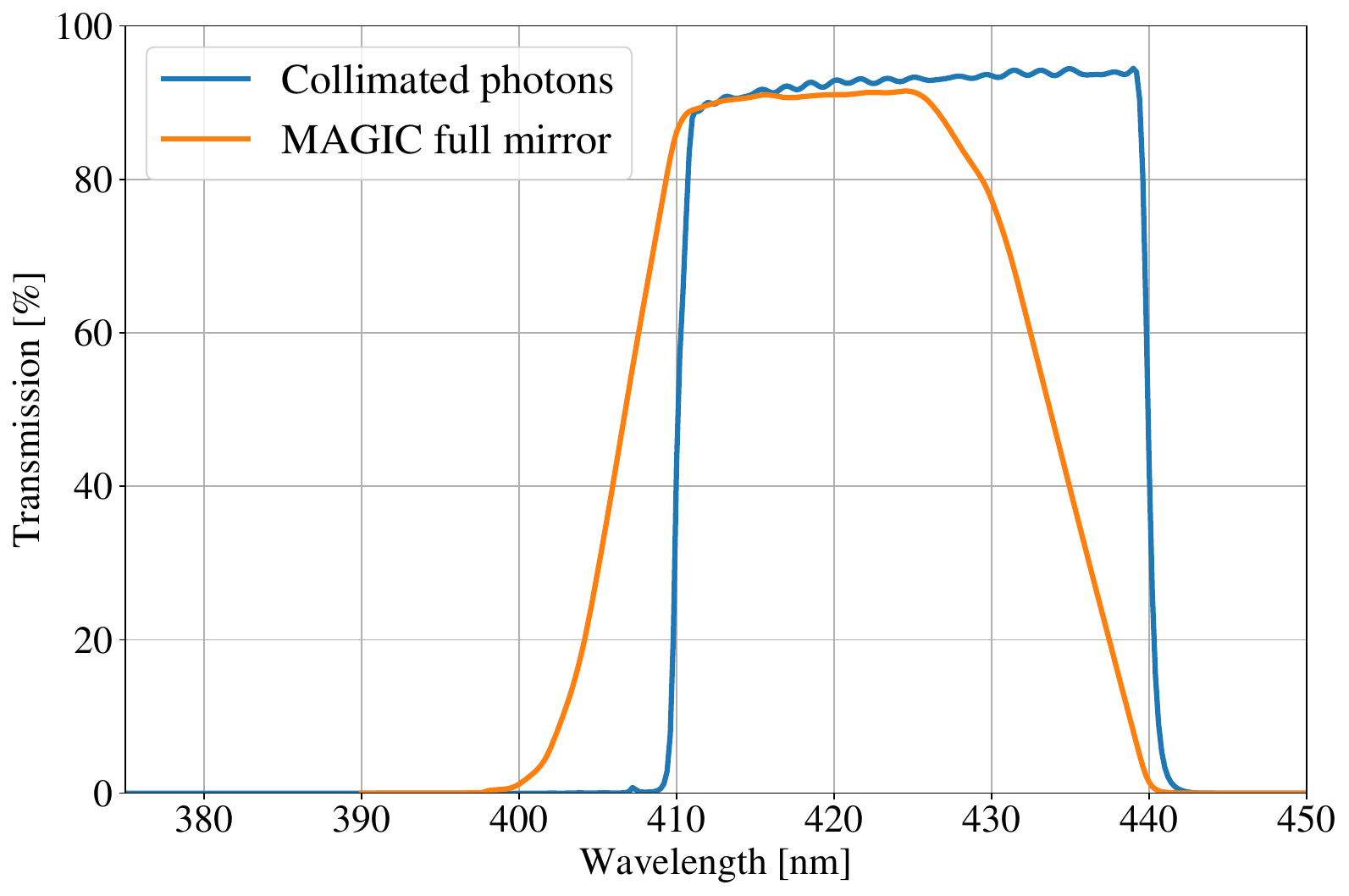}
\end{tabular}
\end{center}
\caption[transmission-curve] 
{ \label{fig:transmission-curve} 
Transmission curve of the Semrock 425-26 nm optical filter, both for a collimated beam (blue) and for an angle distribution as the one expected from light collected by the full MAGIC 17-m diameter reflector (orange). Exclusively the effect of the angle distribution is shown here. Reflectivity and optical performance are expected to modify the transmission, but not the shape of the optical passband.}
\end{figure}

As described by \cite{HB1974}, the S/N of the correlation of telescope signals is insensitive to the width of the optical passband of the detected light. However, a filter spectral response with sharp spectral cutoffs does improve the sensitivity of the measurement. In addition, accounting for MAGIC photo-detection efficiency and gain, we are observing bright stars capable of damaging PMTs after a short exposure time, given the excessive current flowing through the dynode system. For these reasons interferometry observations require installing narrow-band optical filters in front of the PMTs connected to the correlator. We are using interference filters manufactured by Semrock of model 425-26 nm. The spectral transmission curve for incident parallel light is centered at 425 nm and has a FWHM of 26~nm with relatively sharp edges. This shape is strongly modified in our setup because MAGIC has a low f/D ratio (close to 1). We have used the MyLight modeling online tool provided by Semrock\footnote{MyLight online tool for the Semrock 425-26 nm filter accessible \href{https://www.idex-hs.com/store/product-detail/ff01_425_26_25/}{here}.} to calculate the effective spectral transmission curve in a cone of half angle 26.5$^{\circ}$, as shown in Fig.  \ref{fig:transmission-curve}.

\begin{figure}
\begin{center}
\begin{tabular}{c} 
\includegraphics[width=0.95\linewidth, trim={2cm 4cm 8cm 3.5cm},clip]{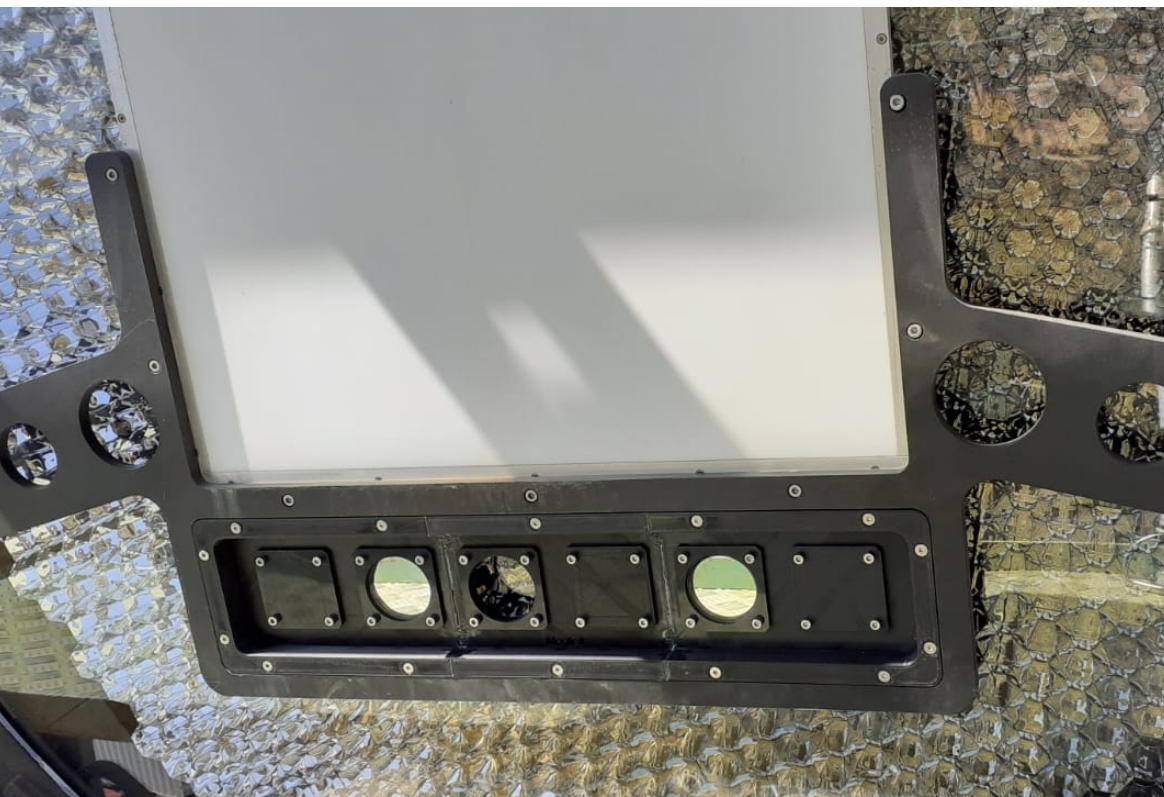}
\end{tabular}
\end{center}
\caption[filter-holder] 
{ \label{fig:filter-holder} 
Filter holder of the MAGIC telescopes below the white target, with room for six filters. At the time the picture was taken only two filters were installed (greenish circles), one of the holes was left open and three holes were closed with a plastic cap. Behind the filter holder and the diffusive white target one can see the hexagonal light concentrators (Winston Cones) right in front of the PMTs.}
\end{figure} 

The filters have a diameter of 50~mm and are held 20~mm in front of MAGIC photo-detectors using a mechanical frame (``filter holder''). This frame piggybacks on an existing mechanical structure that holds a diffusely reflecting white plate used for absolute calibration of the reflectance of the mirror (the so-called ''white target''). This structure is operated remotely and can be deployed in front of the PMTs in a matter of seconds. Fig. \ref{fig:filter-holder} shows a picture of the filter holder under the white target. There is room for six filters in a horizontal line, all centered in front of PMTs that may eventually be equipped for interferometry. The additional filter slots also allow for a simultaneous signal and background monitoring using identical filters during interferometry observations. Each filter is placed and centered in front of the central PMT of the 7-pixel clusters of the MAGIC camera. The pixels used in this work have a $\sim$ 24 cm offset (0.8$^\circ$ on sky) with respect to the center of the camera. 

\subsection{Active mirror control}
\label{sec:amc}

The parabolic-shape reflector of a MAGIC telescope is approximated by using 246 individual mirror facets of spherical shape. The facets are fixed on the reflector according to their radius of curvature, which varies between 34 to 36 m. Eleven groups of such mirrors lead to a very low time spread of synchronous light pulses (< 1 ns) \citep{reflecting_surface, performance_paper}. The light-weight design of the MAGIC reflector, made of reinforced carbon fiber tubes, allows for fast slewing of the telescopes but is subject to deformations of the dish and sagging of the camera. Each mirror facet of 1 m$^2$ is equipped with 2 actuators, correcting for these effects via an Active Mirror Control (AMC) system \citep{amc_icrc}.

The flexibility given by the AMC has been a key asset to the MAGIC intensity interferometry setup. As described by \cite{HB1974}, the sensitivity of  SII strongly relies on the photon detection capabilities of the telescopes. The MAGIC Collaboration devotes significant observational, scheduling,  and analysis efforts for keeping the bending models of the telescopes up-to-date to ensure that the AMC minimizes the effect of structural deformations of the telescopes, maximizing the amount of starlight reaching the right camera pixels \citep{TPOINT}. During interferometry-mode observations, the on-site crew performs a re-focusing of the AMC whenenever the zenith distance changes by at least 5 deg, to ensure the right bending model is applied as stars drift during the observation.

In addition to correcting deformations, the AMC allows for a broad range of configurations, transforming the MAGIC telescopes into a very versatile optical interferometer:
\begin{itemize}
    \item Full-mirror: as described in section \ref{sec:optomechanics}, light from the target star focuses into a single pixel located with a 0.8 deg offset with respect to the camera center. This offset adds negligible spread to synchronous light signals (still < 1 ns). This is the standard observing mode of the MAGIC-SII setup, which focuses starlight into any of the 6 pixels behind the optical filter holders.
    
    \item Chess-board: by focusing half of the mirrors to one interferometry pixel and the other half to another (in a  pattern similar to a chess-board), each MAGIC telescope becomes 2 virtual overlapping telescopes. As described in section \ref{sec:digitizer_correlator}, the GPU correlator is able to handle 4 input signals (computing 6 correlations), so we are able to simultaneously gather long-baseline correlation signals as well as a measurement close to our zero-baseline correlation.
    
    \item Sub-reflectors: as any combination of the mirrors located within MAGIC reflectors may be focused to the interferometry pixels, the MAGIC-SII system has the capability of sampling short-baseline correlations within the 1-17 m range by creating multiple sub-reflectors (of e.g. 3-5 m in diameter). Even if the number of feasible targets to be studied with this configuration is limited (very-bright, very-large stars), the flexibility on the number of sub-reflectors and their relative location allows a broad sampling of the Fourier space of the image (UV coverage). This could be expanded to more complex setups, such as the I3T concept, envisioned by \cite{i3t_gori2021}.
\end{itemize}

\subsection{Digitizer and correlator}
\label{sec:digitizer_correlator}

The correlator hardware and software have been designed to harness the massively parallel nature of state-of-the-art GPUs to process in real time the data captured using two fast digitizer boards Spectrum M4i.4450-x8 PCIe 2.0. 

Each digitizer deals with two channels providing up to 500 MS/s of simultaneous sampling rate with a resolution of 14 bits per sample. Furthermore, the selected digitizers support Remote Direct Memory Access (RDMA), enabling direct data transfers between the digitizer's memory and the GPU's memory. This eliminates the need for intermediate copies, resulting in increased throughput and reduced latency. The two Spectrum boards clocks are synchronized by means of the Star-hub module attached to the carrier card in the correlator chassis. Because using the High Frequency mode introduces a strong flip-flop correlated noise into the data stream, buffered mode (with a resistance of 50 $\Omega$) is used for the digitizer (imposing the $\sim$130 MHz bandwidth described in the previous section).

The correlator is implemented in a computing server with off-the-shelf hardware: two processors (20 cores in total), Solid State Drives (SSDs) for fast access and Hard Drive Disks (HDDs) for longer term storage and tests, and a Nvidia Tesla V100 GPU. The GPU chosen is the PCIe 3.0 x16 model with 5120 cores, 32 GB HBM2 RAM and 14 TFLOPs of single-precision performance.

The correlation code, described in section \ref{sec:online_correlation}, has been used to process data from the two digitizers in real time at 4 GB/s. Alternatively, the code can dump raw data from one digitizer (two channels) to disk in real time for short periods of time, mainly for testing purposes.

\section{Data analysis}
\label{sec:analysis}

We use the prescription described by \cite{magic_2019} to perform intensity interferometry astrophysical measurements with the MAGIC-SII setup, where we define the contrast $c$, proportional to the squared visibility $V^2$:

\begin{equation}
V^2 \propto c = K \frac{\rho({\tau_0}) \sqrt{G_1 G_2}}{\sqrt{DC_1 DC_2}}
\end{equation}
where $K$ is a constant, $\rho$ is the Pearson's correlation at the $\tau_0$ where the signal is expected, $G_i$ are factors applied to correct for changes in the HV of pixel \textit{i}, and $DC_i$ are the DC measured in pixel \textit{i}, which is proportional to the photon flux. This expression assumes the signal is dominated by the stellar flux, and ignores the contribution of the NSB. When NSB currents become a significant fraction of the measured signal, this expression needs to be expanded to account for the contribution of the un-correlated photons:

\begin{equation}
\label{eq:contrast}
c = K \frac{\rho({\tau_0}) \beta \sqrt{G_1 G_2}}{\sqrt{DC(Star)_1 DC(Star)_2}}
\end{equation}
where $\beta$ accounts for the ratio between stellar and NSB photon fluxes:

\begin{equation}
\beta = \sqrt{\frac{(DC(Star)_1 + DC(NSB)_1)(DC(Star)_2 + DC(NSB)_2)}{DC(Star)_1 DC(Star)_2}}\,.
\end{equation}

Squared visibility measurements are calculated using eq. (\ref{eq:contrast}) by performing the following steps: 
\begin{enumerate}
    \item $\rho(\tau)$ computation by the correlator,
    \item calibrate $\rho(\tau)$ with $DC_i$, $G_i$ and $\beta$ factors,
    \item variable time delay $\tau$ correction,
    \item extract $\rho$ at the expected $\tau_0$.
\end{enumerate}

In this section we will describe the implementation developed for the low-level analysis, i.e. the GPU-based online computation of Pearson's correlation $\rho(\tau)$, signal calibration and high-level analysis of square visibility measurements. 

\subsection{GPU-based on-line correlation}
\label{sec:online_correlation}

The main functionality of the correlator software is to compute the Pearson's correlation ($\rho(\tau)$) for all possible pairs from the 4 channels used as a function of the time shift $\tau$ between them over a window wide enough to cover the expected range of delays (in our case, a generous $\pm$ 2048 ns). It also computes the autocorrelation of these channels, resulting in a total of 6 correlations and 4 autocorrelations. To perform these computations efficiently, the software exploits the convolution theorem in frequency domain implemented with Fast-Fourier-Transform (FFT). The correlation for each pair in frequency domain is computed for time frames of a given size (generally $2^{18}$ samples per channel), added in sets of 500 frames and then converted back into $\tau$ domain using the inverse FFT. The resulting correlations, normalized with statistics of the input signal to obtain $\rho(\tau)$, are time stamped and stored in disk together with the mean voltage and standard deviation of each input channel used for the computation.

The software is developed in CUDA C using the Nvidia FFT library\footnote{\url{https://docs.nvidia.com/cuda/cufft/index.html}} for the convolution computation, and the Spectrum CUDA Access for Parallel Processing (SCAPP) SDK\footnote{\url{https://spectrum-instrumentation.com/products/drivers_examples/scapp_cuda_interface.php}} for transferring the data between the digitizers and the GPU in streaming mode. It is divided in three parts:

\begin{enumerate}
\item The initialisation section configures several parameters of the digitizers and the GPU (e.g., number of channels, number of correlations, acquisition rates, input paths and ranges, RDMA, execution time), initialises the data structure in both the CPU and the GPU and creates the thread that writes the resulting correlation to a storage media. The data processing loop then starts using a double buffering scheme for the input and output, which allows for a live-time of the correlator of $\sim$100\%.
\item The data processing loop runs continuously for the programmed duration of data taking (generally, a data run of 5 minutes), until an error occurs or until it is interrupted by the operator. In this loop: 
\begin{enumerate}
    \item The code first checks and retrieves into an input buffer a time frame of a given number of samples from each stream of incoming data from the channels of the digitizers.
    \item Statistics, such as the mean and the standard deviation, are obtained from the input buffer and added to accumulators in an output buffer for each input channel.
    \item Simultaneously, the FFT is computed for the data in the input buffer of each channel.
    \item After this, cross correlations and auto correlations are computed by multiplying the obtained FFT  and complex conjugate for each channel, which are then added to an accumulator in the output buffer.
\end{enumerate}

\item Every 500 iterations in the data processing loop, the output buffer is swapped and a writer thread is awoken. This thread normalizes (using the computed statistics accordingly to the definition of the Pearson's correlation) and saves the accumulated correlation results and any obtained statistics to disk in binary format, sets the accumulators to zero and goes back to sleep. 
\end{enumerate}

In addition to its main functionality, the software can be used to store the raw signal of one digitizer directly to disk. In this mode a sample with a resolution of 14bits is saved every 2 ns for each channel. Due to the size of raw data (1 Gb/s/channel), only data from two channels can be saved simultaneously, typically in runs of 10 s.

\subsection{Coherence calibration and signal extraction}
\label{sec:calibration}

The correlator described in the previous section computes and stores $\rho(\tau)$ in a wide range of time delays ($\pm$ 2048 ns). One $\rho(\tau)$ array is calculated from a fixed integration time, generally lasting $\sim 250$ ms, and stored with their corresponding time tag. As introduced in this section, these signals need to be calibrated using the $DC_i$, $G_i$ and $\beta$ factors, as shown in eq. (\ref{eq:contrast}). As the time $\tau_0$ in which the correlation signal is expected changes along the observation (as the star moves across the sky), $\rho(\tau)$ arrays need to be aligned in time with respect to the expected $\tau_0$ before averaging.

The MAGIC SII setup employs a total of 3 pixels in each telescope. As shown in Fig. \ref{fig:filter-holder}, multiple filter holders are available. Currently, a total of 3 Semrock 425-26 nm optical filters are installed in each telescope: two pixels in each telescope, tabulated in Table \ref{tab:average_pulse}, are used for signal extraction (connected to the correlator) while a third pixel is used exclusively for simultaneous background DC measurements over the same optical passband. DC reports from the telescope slow control system store the DC and HV of each pixel once per second for each camera. DC  conversion factors are calculated from calibration measurements, allowing to calculate $DC(NSB)_i$ at any time, and therefore also $DC(Star)_i$ and $\beta$ terms. Most interferometry observations are performed with the same HV values, but those that were performed with a different configuration are still usable as long as their gains are corrected with the $G_i$ terms, also inferred from calibration observations.

The time delay correction is applied to each $\rho(\tau)$ array stored by the correlator, by subtracting the expected time-delay of the correlation signal $\tau_0$, so that the expected correlation signal will always be located at $\tau \sim 0$. The expected time-delay $\tau_0$ depends on two terms: 1) the specific hardware delay from each channel used in the observation (dominated by the length of the optical fibers) and 2) the time of flight difference between photons detected in each telescope. As current digitizers operate at 500 MHz, arrays are aligned in steps of 2 ns. 

Once the correlation signal of a given source has been calibrated and time-delay corrected (i.e. $\frac{\rho(\tau - \tau_0) \beta \sqrt{G_1 G_2}}{\sqrt{DC(Star)_1 DC(Star)_2}}$) it is collected and accumulated, generally in bins of baseline. The averaging is performed by weighting each array by the measured off-peak variance (using values far from $\tau \sim 0$). As the uncertainty expected of the correlation is inversely proportional to the photon flux measured, a linear average across time would magnify the uncertainty in the correlation when combining observations with different quality (measuring different fluxes, due to cloudiness, pointing direction or hardware issues). After this weighted average, contributions of bad-quality data are minimized according to their expected larger variance, allowing to combine observations with very different data quality with minimal negative impact on signal to noise.

\begin{figure*}
\includegraphics[width=0.8\textwidth]{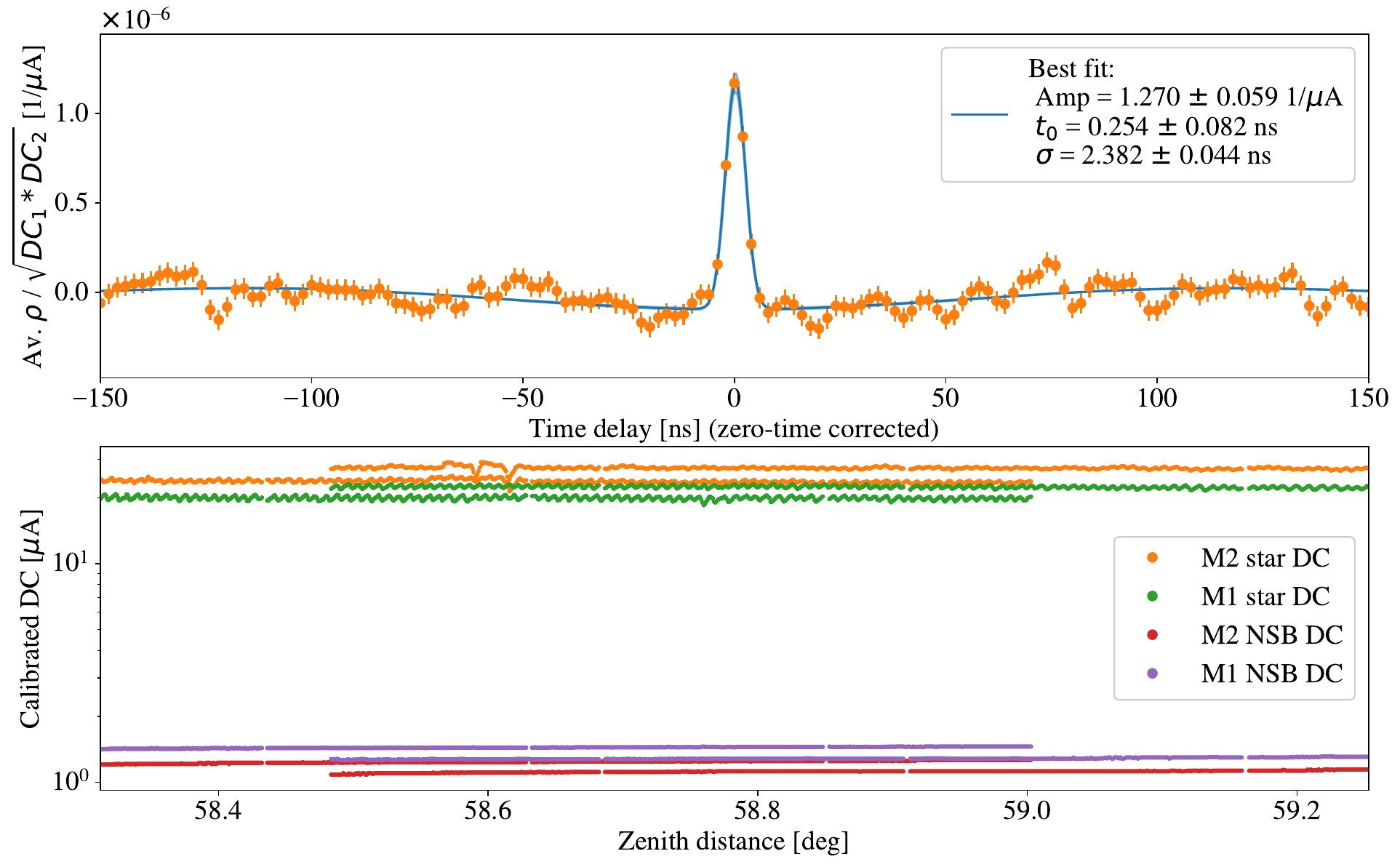}
\caption{Example of calibrated Pearson's correlation signal. \textit{Top}) Weighted average correlation as a function of the zero-time corrected delay ($\tau$ - $\tau_0$) between the two input signals (A-C channels), from $\sim$ 30 min of observation with an average baseline of 52 m pointing to eps CMa (Adhara). The amplitude of the correlation is shown in the legend, using a Gaussian function with a 12 MHz cut high-pass filter applied. Best fit is drawn in blue while 1-$\sigma$ uncertainty region is drawn in orange. \textit{Bottom}) calibrated DC (in $\mu$A) from each MAGIC telescope as a function of the zenith distance of the observation, both for the current associated to the star, as well as the NSB. Different DC values from the same telescope and zenith distance come from observations from different nights. Wind gusts produce relatively large fluctuations in M2 (less protected from wind than M1) while the low-amplitude sawtooth effect seen in both telescopes is due to the azimuth tracking. \label{fig:correlation_example}}
\end{figure*}

After the correlation is averaged in time, the only step necessary to compute $c$, as shown in eq. (\ref{eq:contrast}), is to extract the amplitude of the correlation signal. This is done by fitting a Gaussian function, where the resulting amplitude will be $c$ (see Fig. \ref{fig:correlation_example}). Given the limited bandwidth of the digitizer, data is strongly correlated, and therefore its bi-dimensional correlation matrix is used in the $\chi^2$ minimization to properly weight the residuals. As will be justified in section \ref{sec:systematics}, a high-pass 12 MHz cut filter is applied (both to the data and the Gaussian fitted function) to mitigate a faint correlated noise seen when integrating over long observing periods. The uncertainty of the squared visibility measurement $\Delta c$ is set to the standard deviation of the off-peak correlation in the time-delay range surrounding the expected correlation signal, i.e., from -140 to -40 and from 40 to 140 ns in $\tau - \tau_0$.

MAGIC PMT gains are calibrated during VHE gamma-ray observations by using interleaved calibration pulses \cite{Aleksic2016, performance_upgrade}. By digitizing with the GPU data acquisition (DAQ) these same calibration pulses we are able to store the average response of these pixels to bright short light pulses. Table \ref{tab:average_pulse} shows the width of calibration pulses measured by the different pixels equipped for interferometry observations. In addition, as the GPU DAQ also computes the auto-correlation of each channel, we can ensure starlight observations are providing the expected bandwidth. From the average response to calibration pulses between different pixels we are able to calculate the expected shape of our correlation signal ($\sigma \sim$ 2.2 ns). This value already includes the fixed jitter (0.4 ns)   we expect for full-mirror observations from MAGIC I (the first telescope built in 2004) since mirrors in the reflecting dish  were installed in two layers of facets separated by 60 mm in the optical axis direction. As deviations with respect to a Gaussian are at a few percent level and statistical uncertainties from individual correlation signals will not reach such precision, we use a Gaussian function to extract the amplitude of the correlation signal. The width of the measured pulses is currently limited by the bandwidth of our digitizers.

\begin{table}
\centering
\begin{tabular}{ccccc} 
Telescope & Pixel ID & Label & Input [ns] &  $\sigma$ [ns] \\
\hline
MAGIC-II & 251 & A & $\sim$ 1 & 1.59 $\pm$ 0.03\\
MAGIC-II & 260 & B & $\sim$ 1 & 1.52 $\pm$ 0.03\\
MAGIC-I & 251 & C & $\sim$ 1 & 1.64 $\pm$ 0.03\\
MAGIC-I & 260 & D & $\sim$ 1 & 1.65 $\pm$ 0.03\
\end{tabular}
\caption{Response of individual MAGIC pixels to calibration pulses as measured by the MAGIC-SII GPU-based correlator. Telescope, pixel ID number, input and measured average widths (Gaussian $\sigma$) for each pixel currently equipped for interferometry observations. These are computed by averaging calibration pulses from raw digitized data (\mbox{500 Msa/s}). \label{tab:average_pulse}}
\end{table}

\subsection{Angular diameter analysis}
\label{sec:diameter_analysis}

As described in section \ref{sec:calibration}, correlation data is calibrated, time-delay corrected and bundled to perform the weighted average over the observed time. In this work, data is bundled in uniform steps in baseline, and $V^2$ measurements are extracted from the amplitude of each correlation signal. In the following we will consider either uniform disc (UD) or limb darkened (LD) profiles for the observed stars, i.e. radially symmetric models, therefore baseline $d$ is the only relevant parameter to consider. In the UD scenario, visibility $V$ can be expressed with the Bessel function of the first order $J_1$, as in e.g.  \cite{visibility_modeling}:

\begin{align}\label{eq:Bessel}
	V(d)=2\frac{J_1(\pi d  \theta_{UD}/\lambda)}{\pi  d \theta_{UD}/\lambda},
\end{align}

where $\theta_{UD}$ is the diameter of the uniform disc, $d$ the projected inter-telescope distance (baseline) and $\lambda$ the central wavelength of the optical bandpass of our setup (420~nm, see Fig \ref{fig:transmission-curve}). When including the limb darkening effect, we follow the prescription introduced by \cite{HB_LD_model}, described as

\begin{dmath}
\label{eq:HB_LD}
	{V(d)^2 = \left(\frac{1-u_\lambda}{2} + \frac{u_\lambda}{3}\right)^{-2}} \left( (1-u_\lambda) (\frac{J_1(x_{LD})}{x_{LD}}) + u_{\lambda} \sqrt{\pi/2} \frac{J_{3/2}(x_{LD})}{(x_{LD})^{3/2}}  \right)
\end{dmath},

where $x_{LD} = \pi d  \theta_{LD}/\lambda$, $u_\lambda$ is the limb darkening coefficient and $\theta_{LD}$ is the angular diameter of the limb-darkened star.

\begin{figure*}
\includegraphics[width=0.8\textwidth]{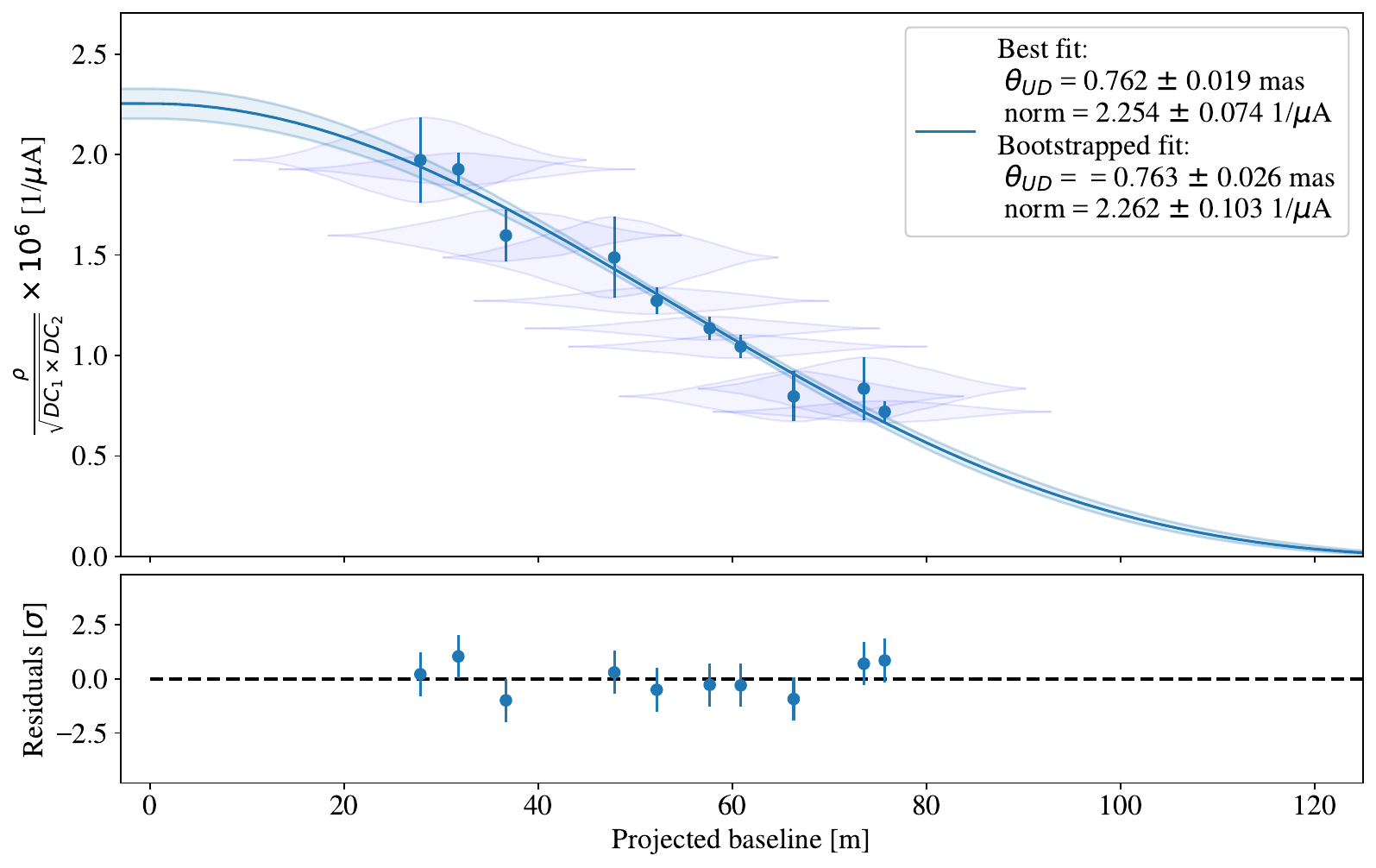}
\caption{Squared visibility vs baseline, showing the uniform-disc stellar diameter measurement of eps CMa using A-C correlation data (M1 pixel 251 correlated with M2 pixel 251). Envelopes surrounding each measurement highlight the true distribution of correlated baselines due to the large size of MAGIC reflecting dishes. Resulting best-fit values of the stellar diameter and normalization, i.e. the zero-baseline correlation $V^2(0)_{A-C}$, are shown in the legend both from the $\chi^2$ minimization as well as a bootstrapping procedure. Bottom figure shows the residuals with respect to the best fit uniform disc model.\label{fig:visibility}}
\end{figure*}

Stellar diameters are measured by fitting eq. (\ref{eq:Bessel}) to the $V^2(d)$ measurements. The statistical uncertainty of fitted parameters, such as $\Delta \theta$ and the zero-baseline correlation $\Delta V(0)$, are considered to be the largest value between the one extracted from the $\chi^2$ minimization method (the value increasing the $\chi^2$ by $\chi_\nu^2$, as in \cite{lmfit}) or the one calculated via bootstrapping. As the size of the MAGIC reflectors is comparable to the distance between the telescopes, eq. (\ref{eq:Bessel}) is evaluated following the true distribution of distances between random points within the two reflectors (see Fig. \ref{fig:visibility}).

Under these assumptions, in order to constrain the diameter of stars with uncertainties down to the few percent level it is necessary to reach such signal to noise both in the $V^2(d)$ and $V^2(0)$. As discussed in \cite{HBbook1974}, the zero-baseline correlation of an intensity interferometer is a constant of the system that mainly depends on the electronic and optical bandwidth of the hardware setup. Given the MAGIC interferometer is only composed by a single pair of telescopes, only stars located in favourable locations in the rotating sky provide the possibility of measuring a wide range of baselines (in the 20 to 87 m range), to reasonably constrain not only $\theta$, but also $V^2(0)$. As the latter is expected to be constant for a given hardware setup, as soon as $V^2(0)$ is properly constrained, its measurement will be used to measure $\theta$ of stars in less favourable positions in the sky, which only allows for the measurement of a very limited range of baselines. For the MAGIC telescopes, the ideal star for constraining $V^2(0)$ is eps CMa (i.e. Adhara). As shown in Fig. \ref{fig:correlation_example} and \ref{fig:visibility}, eps CMa is bright enough to provide strong correlation signals over reasonable observing times, and allows the broad baseline coverage required to strongly constrain both $\theta$ and $V^2(0)$.

As described in \cite{magic_2019}, because the signals from MAGIC are not DC coupled, other parameters may affect the measured $V^2(0)$ value, such as gains of the PMTs, or those associated to the DC measurements. For this reason, from here on we assume independent zero-baseline correlation measurements for each pixel pair: $V^2(0)_{i-j}$. By performing a simultaneous $\chi^2$ minimization to all available observations from eps CMa (with the 4 different channel-pair combinations) and assuming a common $\theta$, we improve the statistical uncertainty of those $V^2(0)_{i-j}$ values with significantly less data (mainly A-D and B-C pairs). Note the UV coverage of all channel pairs is almost identical, so if the source had non-radially-symmetric features, the expected systematic added by this assumption is negligible. The stability of $V^2(0)_{i-j}$ will be discussed in section \ref{sec:systematics}, along with all the sources of systematics evaluated in this work.

\section{Results}
\label{sec:results}

In this section, we summarise some of the results we have achieved with on-sky observations with the MAGIC-SII system, covering its calibration, validation, and new astrophysical measurements of stellar diameters. Since the prototype hardware implementation used in \cite{magic_2019}, the upgraded setup presented here was designed, installed, and commissioned by January 2022. Since then, in the period between January and December 2022, 192 hours of data have been acquired, using a diverse set of observing modes (see section \ref{sec:amc} for a description of the interferometry observing modes): $\sim$ 101 hours of full-mirror observations using pixels 251 (labeled A-C, as in Table \ref{tab:average_pulse}), 28 hours of chess-board observations (correlating the 4 combinations of A, B, C and D channels) and 63.4 hours of full-mirror observations using pixels 260 (labeled B-D).

The targets observed over this period can be broadly classified into two categories: \textit{reference} and \textit{candidate} stars. We consider a target to be a reference star if their diameter has already been directly measured by other instruments over similar wavelengths (\mbox{400-440 nm}). We used the following selection criteria: angular diameter and declination allowing the determination of their stellar diameter with MAGIC baseline, as well as bright enough to detect correlation signals over reasonable observing times. From this selection of stars, we excluded fast rotators, spectroscopic binaries, those having bright close companions ($\Delta B$ < 2.2 and distance below 10 mas) and those showing large variability in magnitude ($\Delta B$ > 1). By measuring the diameter of reference stars we intend to confirm the validity of our analysis and hardware setup. Candidate stars are those not having a direct measurement of their diameter over similar wavelengths, but their predicted size and declination as well as their brightness allow a direct detection of their diameter. Their predicted diameter is extracted from several sources \citep{bourges2017vizier, 2017AJ....153...16S, bonneau2006searchcal,bonneau2011searchcal}.

All data was acquired and analyzed following the steps described in section \ref{sec:analysis}. For simplicity, a fixed step size of 5 m in baseline was used for all sources, even if this binning size may not be optimal for the faintest stars presented. The significant amount of data presented in this work allow us to evaluate the performance of the system, to confirm the validity of analysis  as well as evaluating the scale of the systematics associated to these observations. 

\subsection{Data and analysis consistency}
\label{sec:consistency}

\begin{figure}
\centering
  \includegraphics[width=.49\linewidth]{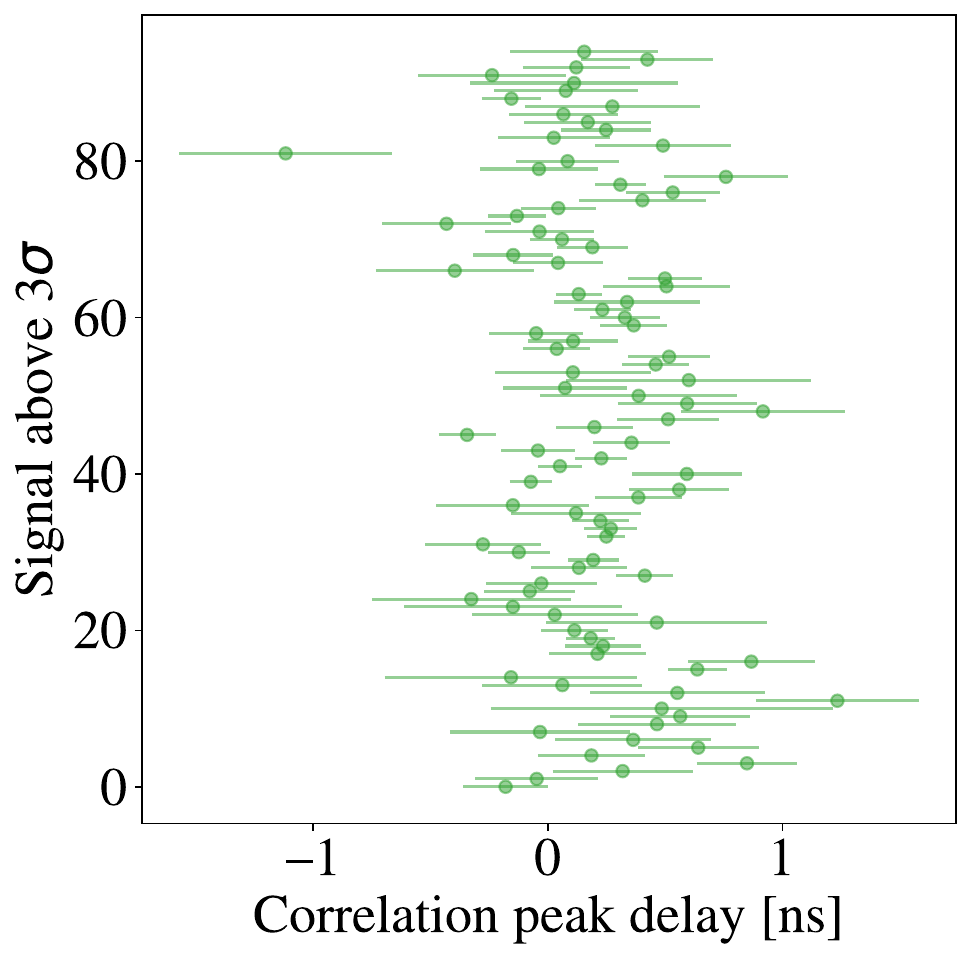}
  \centering
  \includegraphics[width=.49\linewidth]{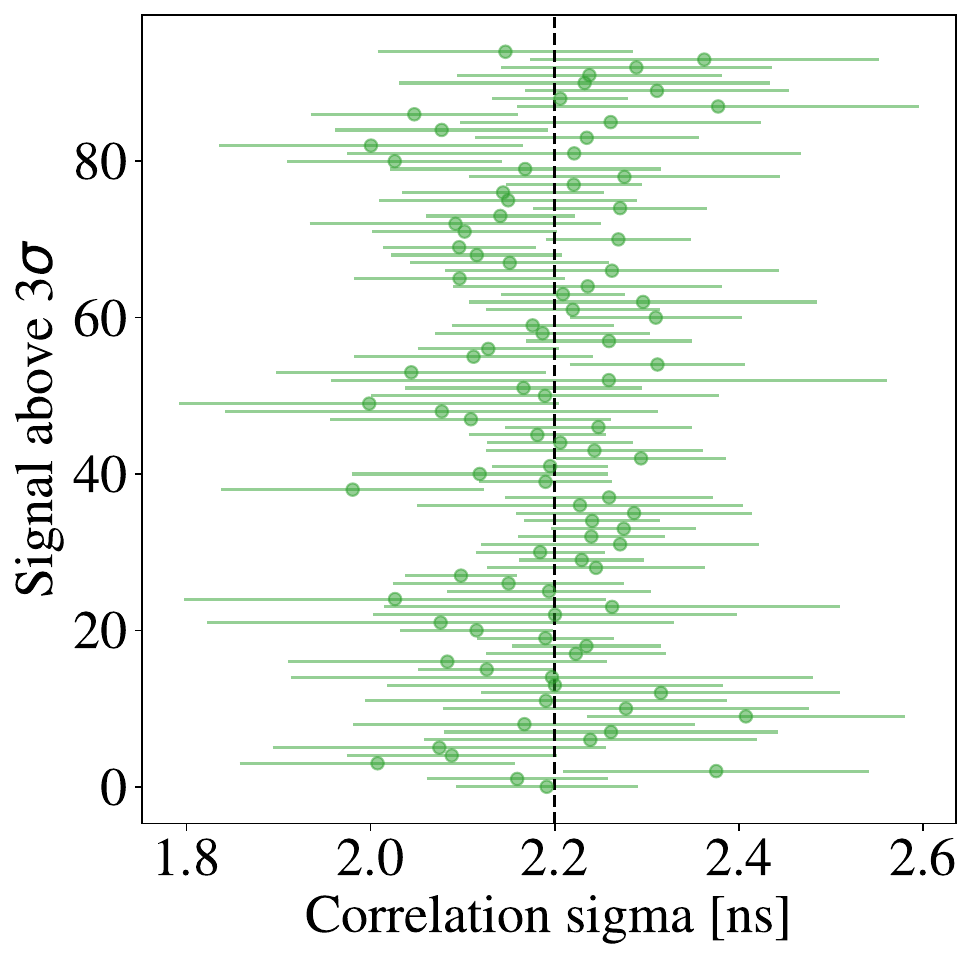}
\caption{Correlation signal delay (left) and width as its Gaussian $\sigma$ (right) for all V$^2$ measurements presented in this work. The dashed vertical line is the expected width of the correlation discussed in section \ref{sec:calibration}.}
\label{fig:correlations}
\end{figure}

Using the analyzed data we can test the scale of the residuals of $\tau-\tau_0$ of the measured correlated signals. We select the correlated signals (each resulting from a set of observations bundled over constant steps in baseline of up to 5 meters) with a signal to noise larger than 3$\sigma$. No matter which pair of channels is used for the correlation, the location of all correlation signals fall within a $\pm$ 1 ns window (as shown in the left panel of Fig. \ref{fig:correlations}, most within $\pm$ 0.5 ns from $\tau-\tau_0$ = 0). The stability of the location of the correlation signal validates the locations of the telescopes assumed in the analysis, the hardware delays assumed for each channel as well as the calculation of photons time of flight as a function of pointing direction.

As discussed in section \ref{sec:photon_transmission}, the expected electronic bandwidth of the system is $\sim$ 125 MHz, dominated by the currently used digitizers. By fitting the width of the measured correlated signals we can test that the performance of the system matches expectations. As introduced in section \ref{sec:online_correlation}, in addition to the 6 possible cross-correlations, the GPU-based correlator is able to store the auto-correlation of each channel. This provides an independent and simultaneous measurement of the electronic bandwidth of each channel (not including all possible effects, such as jitters between different pixels). As shown in the right panel of Fig. \ref{fig:correlations}, the width of all correlated signals is consistent with a Gaussian sigma around 2.2 ns, consistent with the expectations computed both from calibration pulses (see section \ref{sec:calibration}) as well as from measured autocorrelations. This  excludes the presence of strong jitters between pixels, and shows that all pixels equipped for SII observations show very comparable performance.

\begin{figure}
\centering
\includegraphics[width=0.8\columnwidth]{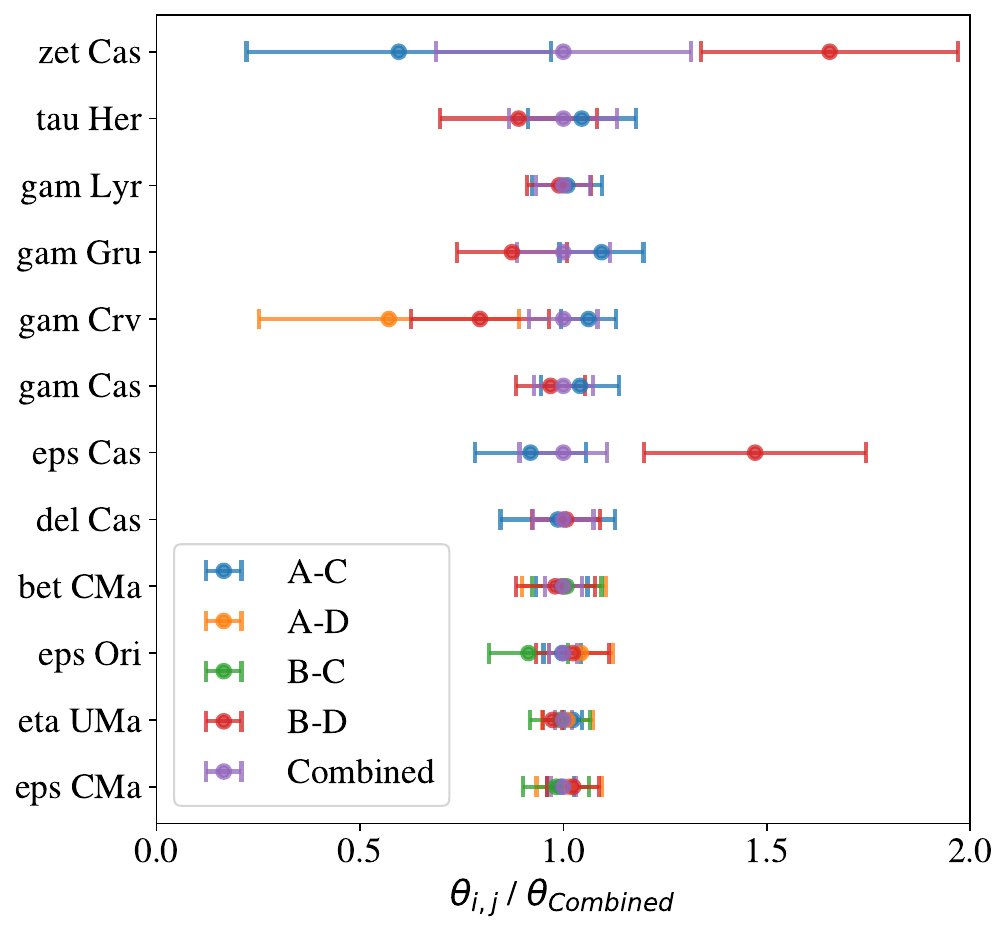}
\caption{Channel-pair-wise stellar diameter measurements relative to the combined reconstructed value. Only measurements of stars having at least two valid measurements with different channel pairs are shown.\label{fig:channel-wise_diameters}}
\end{figure}

The last consistency check performed to evaluate the validity of our analysis is to reconstruct all stellar diameters treating each channel pair as an independent dataset. As introduced in this chapter, the largest fraction of data employs A-C and B-D channels (101 and 63 h respectively) but an additional 28 h were taken in chessboard mode, in which all four correlations are possible (A-C, A-D, B-C, and B-D) although with half of the mirror area. In order to perform these channel-wise stellar diameter measurements, $V^2_{i,j}(0)$ are computed independently for each channel by performing a simultaneous fit of all sources with available $V^2_{i,j}$. As shown in Fig. \ref{fig:channel-wise_diameters}, stellar diameter measurements over the different correlation pairs show remarkable agreement, down to uncertainties in the few-percent level for at least two stars.

\subsection{Stellar diameter measurements}

\begin{figure}
\centering
\includegraphics[width=0.9\columnwidth]{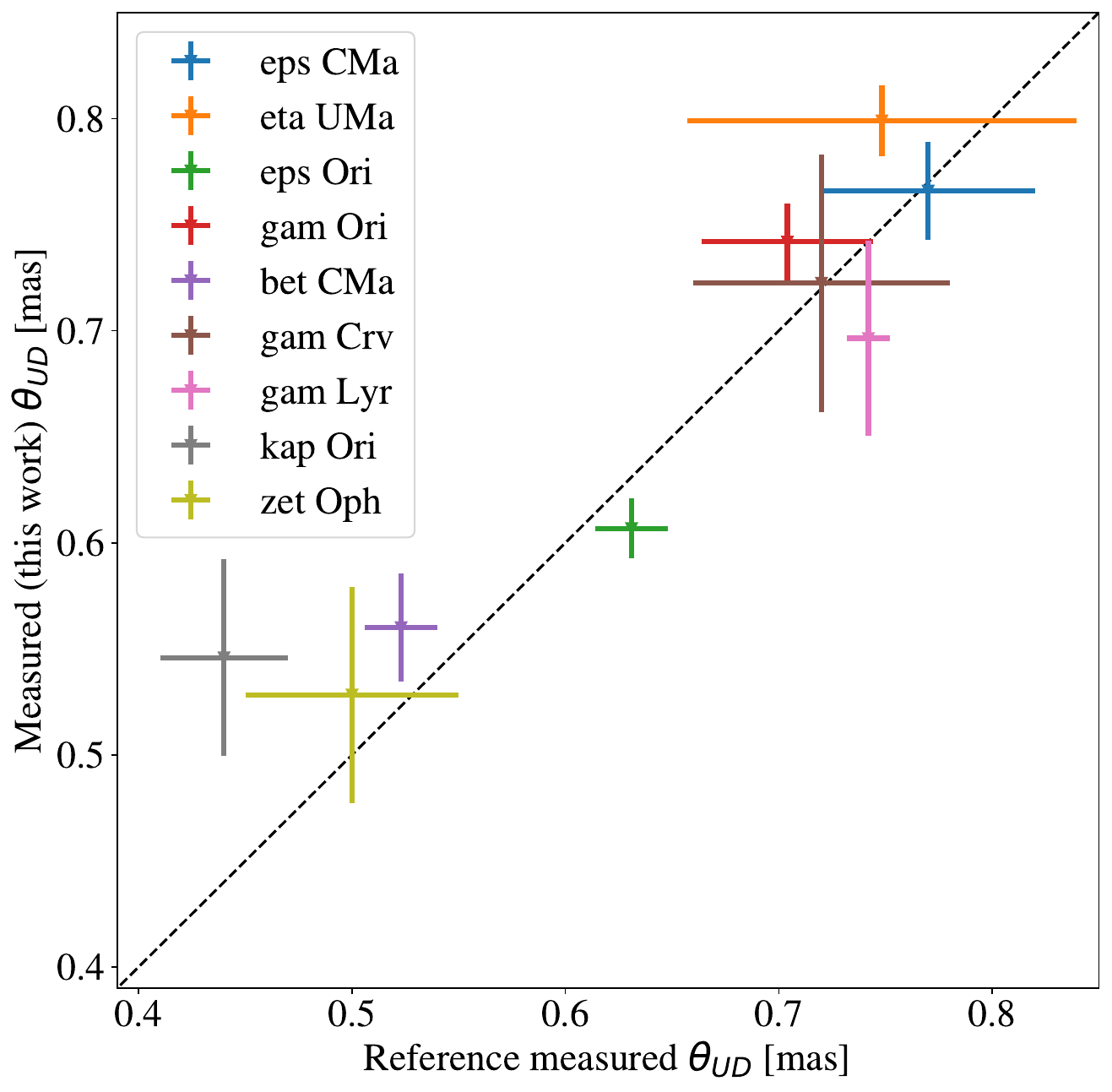}
\caption{Direct measurement of the angular stellar diameter of 9 reference stars, assuming a uniform disk profile, as a function of their reference diameter, also coming from a direct measurement over similar wavelengths. See Table \ref{tab:calib_stars} for their $\theta_{UD}$, $\theta_{LD}$, assumed physical parameters and the source of the reference measurement. Dashed black line shows where the reference equals the measured diameter values.\label{fig:reference_stars}}
\end{figure}

Here we report stellar diameter measurements performed between January and December 2022. As introduced in section \ref{sec:results}, only a few of the stars MAGIC-SII observed have a previous direct measurement of their angular diameter over similar blue wavelengths, allowing a one-to-one comparison of the measurements. The angular diameter measured by MAGIC as a function of the one measured by other instruments is shown in Fig. \ref{fig:reference_stars}. The measurements performed by the MAGIC-SII system are consistent with previous measurements, both from the NSII, VERITAS and CHARA \citep{HB1974, VERITAS, chara_eta_uma}. All the stellar diameter measurements and physical parameters used to determine linear limb darkening coefficients ($u_\lambda$) together with the reference values we compare with are listed in Table \ref{tab:calib_stars}.

\begin{table*}
\centering
\begin{tabular}{ccccccccccc} 
Name	&	HD 	&	Sp. Type	&	B	&	Reference $\theta_{UD}$	&	Source	&	Teff	&	log(g)	&	$u_\lambda$	&	Measured $\theta_{UD}$	&	Measured $\theta_{LD}$	\\
	&		&		&	(mag)	&	(mas)	&		&	K	&	cm/s²	&		&	$\pm$ Stat $\pm$ Syst (mas)	& $\pm$ Stat $\pm$ Syst (mas) \\
\hline                     
eps CMa	&	52089	&	B1.5II	&	1.29	&	0.77 $\pm$ 0.05	&	NSII	&	24750	&	3.65	&	0.364	&	0.768 $\pm$ 0.023 $\pm$ 0.019 &	0.792 $\pm$ 0.023 $\pm$ 0.020	\\
gam Ori	&	35468	&	B2V	&	1.42	&	0.704 $\pm$ 0.04	&	NSII	&	22570	&	3.72	&	0.368	&	0.742 $\pm$ 0.017 $\pm$ 0.010 &	0.765 $\pm$ 0.017 $\pm$ 0.010	\\
eps Ori	&	37128	&	B0Ia	&	1.51	&	0.631 $\pm$ 0.017	&	VERITAS	&	27000	&	2.85	&	0.465	&	0.606 $\pm$ 0.020 $\pm$ 0.018 &	0.630 $\pm$ 0.022 $\pm$ 0.019 \\
eta UMa	&	120315	&	 B3V	&	1.67	&	0.818 $\pm$ 0.06	&	CHARA	&	17200	&	3.78	&	0.403	&	0.800 $\pm$ 0.017 $\pm$ 0.011 &	0.828 $\pm$ 0.018 $\pm$ 0.011\\
bet CMa	&	44743	&	B1II-III	&	1.73	&	0.523 $\pm$ 0.017	&	VERITAS	&	26630	&	3.89	&	0.348	&   0.560 $\pm$ 0.026 $\pm$ 0.023 &	0.576 $\pm$ 0.026 $\pm$ 0.023	\\
kap Ori	&	38771	&	B0.5Ia	&	1.88	&	0.44 $\pm$ 0.03	&	NSII	&	26500	&	2.70	&	0.488	&	0.545 $\pm$ 0.046 $\pm$ 0.030 &	0.568 $\pm$ 0.048 $\pm$ 0.031	\\
gam Crv	&	106625	&	B8III	&	2.47	&	0.72 $\pm$ 0.06	&	NSII	&	12360	&	3.50	&	0.479	&	0.722 $\pm$ 0.061 $\pm$ 0.024 &	0.753 $\pm$ 0.064 $\pm$ 0.025	\\
zet Oph & 149757 & O9.2IVnn & 2.58 & 0.50 $\pm$ 0.05 & NSII & 32000 & 3.85 & 0.334 & 0.524 $\pm$ 0.052 $\pm$ 0.015 & 0.539 $\pm$ 0.053 $\pm$ 0.015\\
gam Lyr & 176437 & B9III & 3.2 & 0.742 $\pm$ 0.010 & CHARA & 10080 & 3.50 & 0.556 & 0.696 $\pm$ 0.046 $\pm$ 0.016 & 0.733 $\pm$ 0.049 $\pm$ 0.017 \\

\end{tabular}
\caption{Table of reference stars sorted by B magnitude. Spectral types and B magnitudes are from SIMBAD. The reference angular diameters are from the NSII \protect\citep{HB1974}, CHARA \protect\citep{chara_eta_uma, chara_gam_lyr} and VERITAS \protect\citep{VERITAS}). Physical parameters (effective temperatures and surface gravity) extracted from: \protect\cite{physical_pars_1, physical_pars_2, physical_pars_3}. Limb darkening linear coefficient $u_{\lambda}$ interpolated from \protect\cite{ld_linear_coeffs} as done in \protect\cite{VERITAS}. Measurements performed with the MAGIC-SII system are shown, assuming both uniform disc and limb darkened profiles, showing statistical uncertainties and maximum expected systematic deviation. \label{tab:calib_stars}}
\end{table*}

\begin{figure}
\centering
\includegraphics[width=0.85\columnwidth]{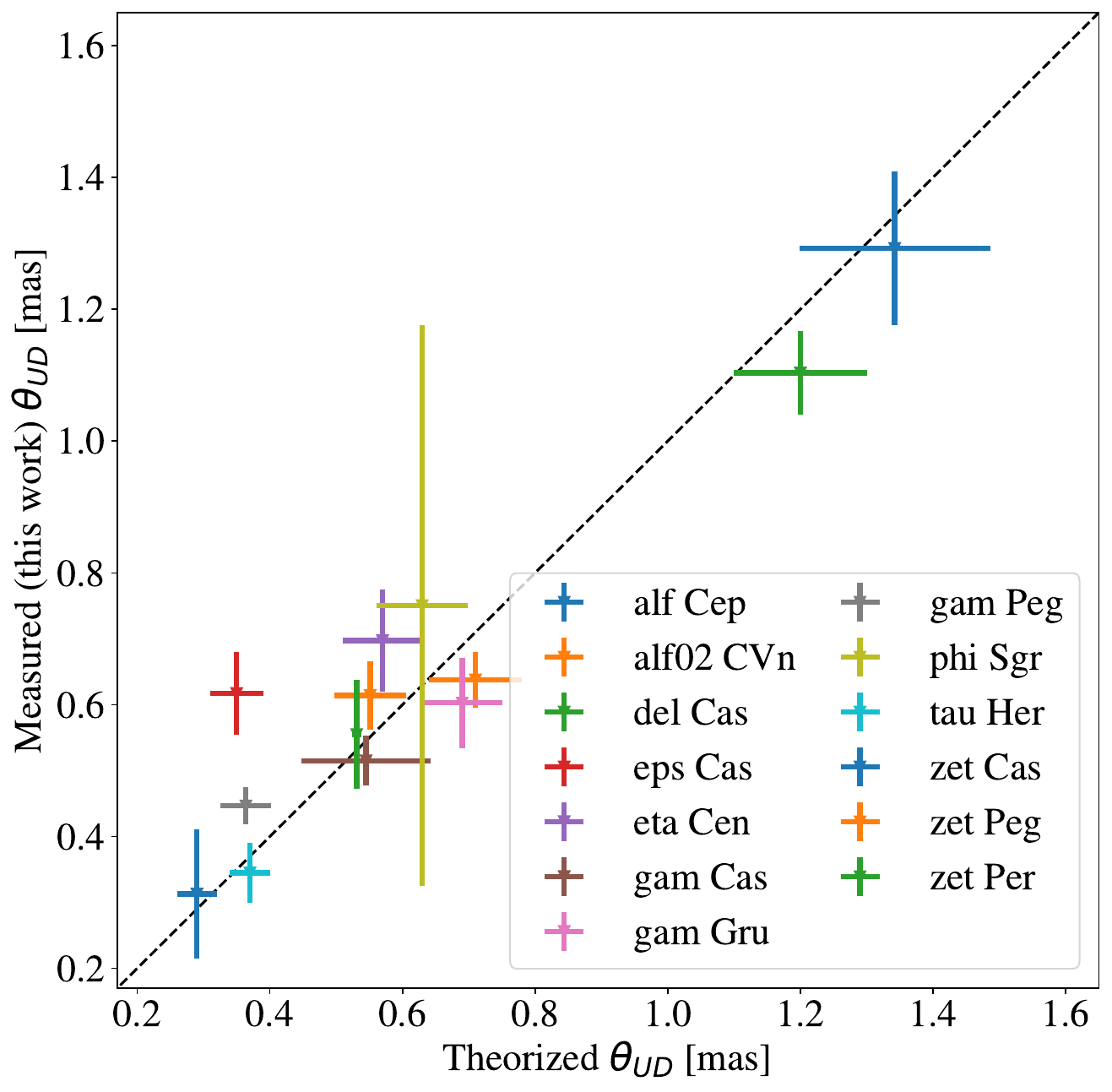}
\caption{Direct measurement of angular stellar diameter of 13 stars newly measured by MAGIC, assuming a uniform disk profile, as a function of their expected angular diameter from \protect \cite{bourges2017vizier}. See Table \ref{tab:candid_stars} for their $\theta_{UD}$, $\theta_{LD}$ and assumed physical parameters. Note the large uncertainty associated to phi Sgr is due to the very small observation time acquired (15 min).\label{fig:candidate_stars}}
\end{figure}

In addition to the reference stars we observed, we selected 13 stars for which their diameters have not been directly measured before in our bandwidth (400-440nm). Table \ref{tab:candid_stars} lists MAGIC new stellar diameter measurements (both using UD and LD models) together with the physical parameters used to determine $u_\lambda$. They are mostly early-type stars between 2.07 and 3.73 magnitudes in B and estimated angular diameters between 0.3 and 1.3 mas. As shown in Fig. \ref{fig:candidate_stars}, MAGIC measurements are again nicely consistent with expectations. A significant fraction of these sources are known fast rotators, and due to the equatorial bulge produced by the centrifugal force they may deviate significantly from the radially symmetric models assumed here. In future works, the MAGIC Collaboration will release interferometric observations following community standards such as OIFITS data products \citep{oifits}.

\begin{table*}
\centering
\begin{tabular}{ccccccccccc} 
Name & HD & Sp. Type & B & Estimated $\theta_{UD}$ & Source & Teff & log(g) & $u_\lambda$ & Measured $\theta_{UD}$ & Measured $\theta_{LD}$ \\
 &  &  & (mag) & (mas) &  & K & cm/s² &  & (mas) & (mas) \\
\hline                     

eta Cen & 127972 &  B2Ve & 2.12 & 0.570 $\pm$ 0.059 & JSDC & 22336 & 4.09 & 0.360 & 0.700 $\pm$ 0.069 $\pm$ 0.021 & 0.723 $\pm$ 0.074 $\pm$ 0.021\\
gam Cas & 5394 & B0.5IVpe & 2.29 & 0.545 $\pm$ 0.098 & Swihart & 27990 & 3.32 & 0.398 & 0.515 $\pm$ 0.038 $\pm$ 0.023 & 0.532 $\pm$ 0.039 $\pm$ 0.023\\
gam Peg & 886 & B2IV & 2.61 & 0.364 $\pm$ 0.037 & JSDC & 37079 & 4.08 & 0.297 & 0.445 $\pm$ 0.029 $\pm$ 0.019 &	0.459 $\pm$ 0.029 $\pm$ 0.020	\\
 &  &  &  & 0.471 $\pm$ 0.019 & Swihart &  &  &  &  &  \\
alf Cep & 203280 &  A8Vn &  2.68 & 1.34 $\pm$ 0.14 & JSDC & 7217 & 3.69 & 0.655 & 1.29 $\pm$ 0.12 $\pm$ 0.04 & 1.37 $\pm$ 0.12 $\pm$ 0.04 \\
alf02 CVn & 112412 & A0VpSiEu & 2.76 & 0.706 $\pm$ 0.067 & JSDC & 9164 & 4.50 & 0.579 & 0.634 $\pm$ 0.042 $\pm$ 0.011 &	0.670 $\pm$ 0.042	$\pm$ 0.011 \\
 &  &  &  & 0.59 $\pm$ 0.13 & Swihart &  &  &  &  &  \\
del Cas & 8538 & A5IV & 2.81 & 1.20 $\pm$ 0.12 & JSDC & 7980 & 3.75 & 0.669 & 1.102 $\pm$ 0.078 $\pm$ 0.035 &	1.172 $\pm$ 0.087 $\pm$ 0.037	\\
 &  &  &  & 1.09 $\pm$ 0.10 & Swihart &  &  &  &  &  \\
gam Gru & 207971 & B8IV-Vs & 2.89 & 0.699 $\pm$ 0.057 & JSDC & 12380 & 3.60 & 0.472 & 0.602 $\pm$ 0.069 $\pm$ 0.034 &	0.627 $\pm$ 0.072 $\pm$ 0.035	\\
zet Per & 24398 & B1Ib & 2.97 & 0.606 $\pm$ 0.058 & SearchCal & 28444 & 3.17 & 0.415 & 0.56 $\pm$ 0.10 $\pm$ 0.03 & 0.58 $\pm$ 0.11 $\pm$ 0.03	\\
phi Sgr & 173300 & B8III & 3.05 & 0.630 $\pm$ 0.069 & JSDC & 12550 & 3.64 & 0.471 & 0.77 $\pm$ 0.43 $\pm$ 0.06 &	0.79 $\pm$ 0.46	$\pm$ 0.06 \\
eps Cas & 11415 &  B3Vp\_sh & 3.22 & 0.350 $\pm$ 0.038 & JSDC & 14806 & 3.55 & 0.429 & 0.617 $\pm$ 0.068 $\pm$ 0.037 & 0.638 $\pm$ 0.069 $\pm$ 0.038 \\
zet Peg & 214923 &  B8V & 3.33 & 0.551 $\pm$ 0.054 & JSDC & 11065 & 4.00 & 0.514 & 0.610 $\pm$ 0.049 $\pm$ 0.013 & 0.639 $\pm$ 0.054 $\pm$ 0.014 \\
zet Cas & 3360 & B2IV & 3.47 & 0.288 $\pm$ 0.031 & JSDC & 16857 & 3.63 & 0.408 & 0.31 $\pm$ 0.10 $\pm$ 0.05 & 0.32 $\pm$ 0.10  $\pm$ 0.05 \\
 &  &  &  & 0.319 $\pm$ 0.013 & Swihart &  &  &  &  &  \\
tau Her & 147394 & B5IV & 3.73 & 0.373 $\pm$ 0.032 & JSDC & 14824 & 3.65 & 0.427 & 0.345 $\pm$ 0.046 $\pm$ 0.035 &	0.357 $\pm$ 0.048 $\pm$ 0.036	\\

\end{tabular}
\caption{Table of stars with newly measured angular diameters by MAGIC, sorted by B magnitude. Spectral types and B magnitudes are from SIMBAD. The estimated angular diameters are from the JSDC v2 catalog, the Swihart catalog and the SearchCal tool \protect\citep{bourges2017vizier, 2017AJ....153...16S, bonneau2006searchcal,bonneau2011searchcal}. Physical parameters (effective temperatures and surface gravity) extracted from: \protect\cite{physical_pars_1, physical_pars_2, physical_pars_3, physical_pars_4, physical_pars_5}. Limb darkening linear coefficient $u_{\lambda}$ interpolated from \protect\cite{ld_linear_coeffs} as done in \protect\cite{VERITAS}. Measurements performed with the MAGIC-SII system are shown, assuming both uniform disc and limb darkened profiles, showing statistical uncertainties and maximum expected systematic deviation. Note the large uncertainty associated to phi Sgr is due to the very small observation time acquired (15 min).\label{tab:candid_stars}}
\end{table*}

\subsection{Sensitivity evaluation}

As introduced in \cite{magic_2019}, from equation 5.17 in \cite{HBbook1974} we are able to calculate the S/N we expect for a given correlation signal. The equation, expanded to also account for the effect of the NSB:

\begin{dmath}
\label{eq:sensitivity}
S/N = A\cdot \alpha(\lambda_0) \cdot q(\lambda_0) \cdot n(\lambda_0) \cdot |V|^2(\lambda_0, d) \cdot \sqrt{b_\nu} \cdot F^{-1} \cdot \sqrt{T/2} \cdot (1+\beta)^{-1} \cdot \sigma 
\end{dmath}
where $A$ is the mirror area, $\alpha(\lambda_0)$ is the quantum efficiency at the peak of the optical passband, $q(\lambda_0)$ the optical efficiency of the rest of the system, $n(\lambda_0)$ is the stellar differential photon flux, $|V|^2(\lambda_0, d)$ is the squared visibility at the observed wavelength and baseline, $b_\nu$ the effective cross-correlation electrical bandwidth, $F$ the excess noise factor of the PMTs, $T$ the observation time, $\beta$ the average background to starlight ratio and $\sigma$ the normalized spectral distribution of the optical passband (see eq. 5.6 in \cite{HBbook1974}). More precisely, parameters $A$, $\alpha$, $q(\lambda_0)$, $F$, $\beta$ and $\sigma$ must be understood as the geometric mean between the two telescopes. These parameters, first estimated by \cite{magic_2019}, have been updated and are shown in Table \ref{tab:sensitivity}.

\begin{figure}
\begin{center}
\includegraphics[width=\columnwidth]{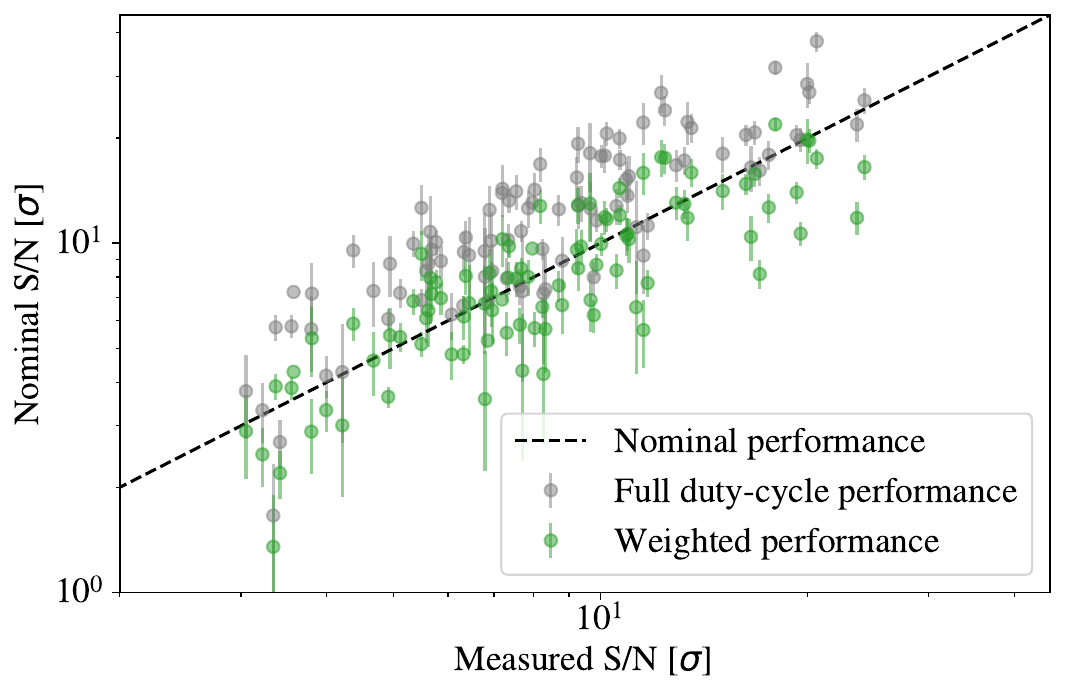}
\end{center}
\caption[signal_to_noise] 
{ \label{fig:signal_to_noise} 
Comparison between the measured and expected signal to noise of each correlation signal. Grey points use the total observing time of each observation to compute the expected signal to noise, while green points take into account the weighting applied (described in section \ref{sec:analysis}) to minimize the impact of low-quality data (observations with significantly lower photon flux).}
\end{figure}

To test if the acquired data matches the sensitivity expected all data presented in this work was compared with the expected signal to noise inferred from eq. (\ref{eq:sensitivity}). As discussed in section \ref{sec:analysis}, no quality cuts are applied to the data because a weighting is applied in the signal averaging step to account for the variable quality conditions of the observations. As shown in Fig. \ref{fig:signal_to_noise}, once the signal weighting is taken into account when computing the expected signal to noise (by calculating a weighted observing time per correlation signal), the observed sensitivity matches the expectations.

\begin{figure}
\includegraphics[width=\columnwidth]{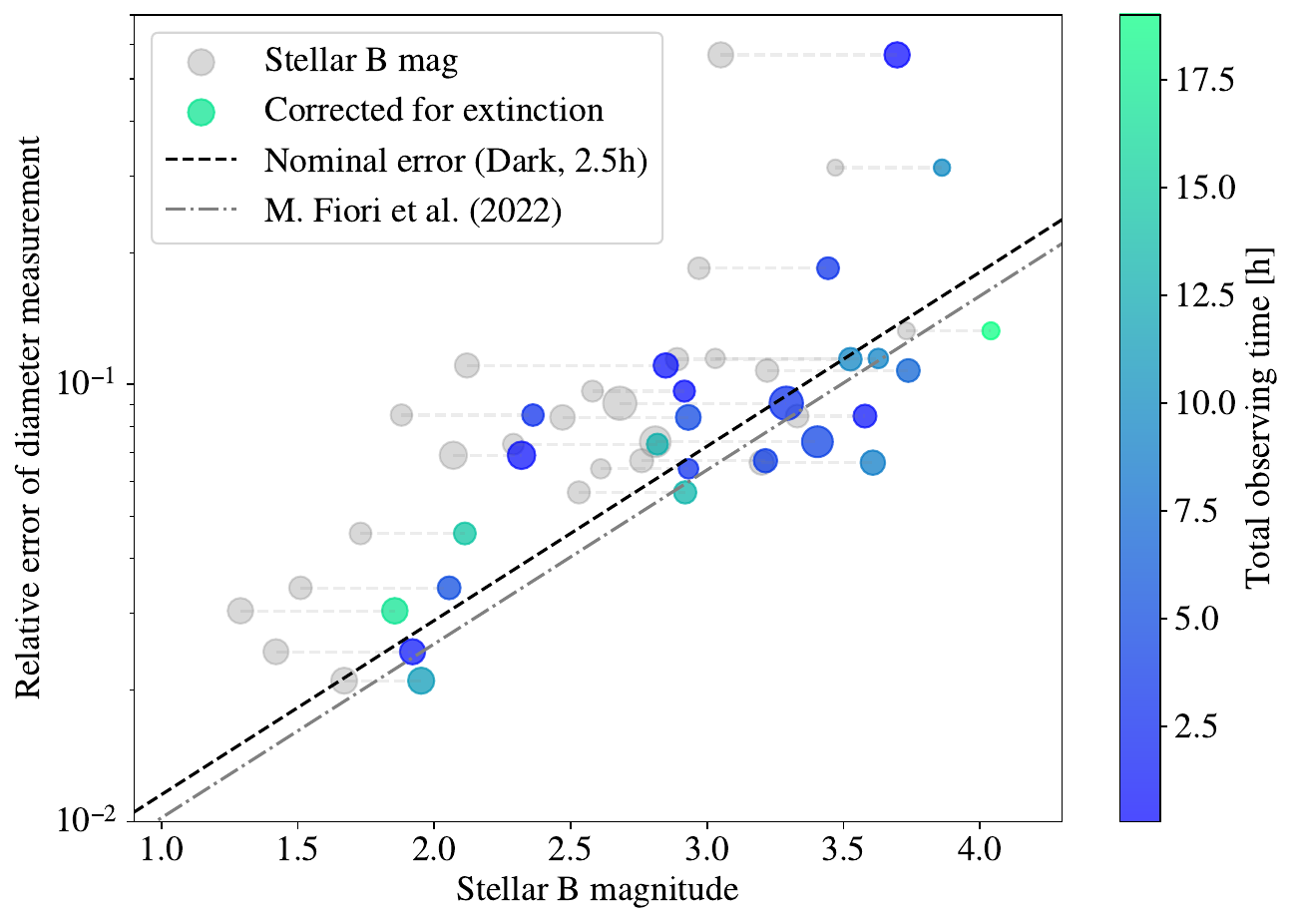}
\caption{Relative uncertainty of measured stellar diameters as a function of their B magnitude. The diameter of each point is proportional to the measured stellar diameter, and the colour scale shows the total observing time used (no quality cuts applied). The nominal expected error as calculated by \protect\cite{spie2022} is shown as a dashed line, and as calculated by \protect\cite{fiori} as a dashed-doted line. Grey points use the average B magnitude, while coloured ones are corrected for the atmospheric extinction affecting the observations. \label{fig:RelativeErrorDiameters}}
\end{figure}

We can also test the resulting relative uncertainty of the angular diameter measurements reported in this work, and compare them with the expected uncertainties. The resulting relative uncertainty of these measurements are shown as a function of stellar B magnitude in Fig. \ref{fig:RelativeErrorDiameters},
before (grey circles) and after (coloured circles) correcting stellar B magnitude with the average atmospheric absorption. Note multiple effects may deviate our measurements from nominal performance: the different total exposure times acquired for each star, the different UV coverage acquired for different stars (MAGIC has only two telescopes at a fixed location) and the very different night-sky  brightness during the observation of each star. However, the achieved relative uncertainty of the measured diameters are in good agreement with the expected trend. 

\begin{table}
\centering
\begin{tabular}{l|r} 
Sensitivity term & Value \\
\hline
Mirror area & 236 m$^2$ \\
Photo-detector QE ($\alpha (\lambda_0)$) & 0.295 \\
Optical efficiency ($q (\lambda_0)$) & 0.304\\
Electronic bandwidth ($b_\nu$) & 125 MHz\\
Normalized spectral distribution ($\sigma$) & 0.87 \\
Noise factor (F) & 1.15\\
\end{tabular}
\caption{Estimated performance parameters of the MAGIC-SII setup. \label{tab:sensitivity}}
\end{table}

\subsection{Systematics evaluation}
\label{sec:systematics}

The systematic uncertainties associated with the analysis of data taken from VHE gamma-ray sources, i.e. the main scientific purpose of the MAGIC telescopes, have been studied in detail in previous works \citep{performance_paper, performance_upgrade}. SII observations are not affected by most of the systematics generally associated to this technique: Uncertainties associated to the energy scale (number of photoelectrons detected), such as quantum efficiency, gain evolution effects due to the VCSELs, mispointing, untracked atmospheric transmission or light collection efficiency of the system only affect the S/N of the correlated signal, and not the measured Pearson's correlation. 

Atmospheric turbulence is expected to add tiny differences in the actual path length of individual photons \citep{HBbook1974}. This turbulence would ultimately become the limiting factor for $b_\nu$ of an optical interferometer (in the case of perfect mirrors and photodetectors with ideal timing). The scale of these differences is expected to be below 30 ps, as calculated by \cite{atmosphere_times}, and therefore the effect is negligible for MAGIC-SII at the  bandwidth scale of the current system. As expected, no significant broadening of MAGIC-SII correlation signals has been detected over high-zenith observations (up to 70$^\circ$ in zenith distance).

Effects modifying the expected zero-baseline correlation (ZBC) of the system, or introducing a miscalibration in the visibility computation will enter the analysis as a systematic. The main sources of systematics that have been identified, summarised in Table \ref{tab:systematics}, are the following:

\subparagraph{Stability of electronic bandwidth:} As described in \cite{HBbook1974}, an untracked evolution of the electronic bandwidth of the system would lead to a modification of the ZBC, which needs to be stable for a reliable analysis. The single-photoelectron response of PMTs is considered stable, but as described in section \ref{sec:photon_transmission}, the signal transmission through VCSel's could perhaps add a time-dependent variation to the bandwidth. As described in section \ref{sec:digitizer_correlator}, the flexibility of the GPU correlator allows a simultaneous measurement of the cross-correlation between input signals and their auto-correlation, which translates into a simultaneous measurement of the electronic bandwidth of each channel. Auto-correlation measurements show the electronic bandwidth of the system is extremely stable over yearly timescales, with variations well below the 0.5\% level (largest rare deviations on the 1\% level, which can be easily identified).

\subparagraph{Optical bandwidth:} The optical bandpass, i.e. the resulting distribution in wavelength of the surviving photons after the narrow-pass optical filter in front of MAGIC PMTs, is the other parameter that would dominate a ZBC evolution. As described in section \ref{sec:optomechanics}, the optical bandwidth of the system is dominated by the wide angle of incidence of photons reaching the filter. MAGIC AMC tracks the number and location of mirrors properly focused (which may vary between different observations). Dedicated toy MC simulations were employed to estimate the impact of missing mirrors over $q(\lambda_0)$: 1) when missing mirrors are randomly distributed on the dish or 2) when removing 4\% of the outer layer of mirrors the effect is still imperceptible; 3) when removing 4\% of mirrors in the central part of the dish the effect is perceptible, but extremely small (< 1\% level); 4) Only when removing a large fraction (more than 20 mirror facets) from the center of the dish the effect is larger than 1\%, situation which never happened in the data presented here. Another effect contributing into the systematics, described by \cite{HBbook1974}, is the mirror deformation as a function of pointing elevation. Dish deformation biases the amount of mirrors actually illuminating the pixel, and may modify the true mirror area (and therefore photon incidence angle distribution on the narrow-band optical filter). In the case of the NSII, telescopes were equipped with optics collimating the beam, improving the performance of the optical filter but increasing the effect of this systematic. MAGIC, on the other hand, is equipped with AMC, which strongly reduces this effect, making it second order for full-mirror observations, as the location of unfocused mirrors is not dominated by the dish bending. 

\subparagraph{Gain evolution of DC readout:} MAGIC slow control measures the DC of all camera pixels to track their illumination level and ensure their safety. As described in section \ref{sec:calibration}, these DC measurements from MAGIC slow control are used to transform Pearson's correlation into visibility measurements. Therefore, any un-tracked gain variation within the DC ADC branch (PMT or DC ADC gains) will modify visibility measurements, and therefore will contribute into the systematics. The authors have identified several effects that could lead to gain variations in the DC branch: 

\begin{enumerate}
    \item temperature variations between summer and winter. Even if temperature within MAGIC cameras are controlled via cooling, the average temperature of the camera differs between seasons. Devoted lab measurements confirmed that the temperature compensation of the 12-bit ADC of the DC monitoring branch provides identical current measurements down to the part-per-million level.
    \item PMT gains evolution over bright light source exposure. Lab test with pixel clusters identical to those used in the M2 camera indicate that when a PMT with a HV set in dark conditions is suddenly exposed to a bright light source (equivalent to a jump in current of $\sim$ 20 uA) the pixel gain undergoes a recovery period, asymptotically reducing the measured DC. From the controlled tests performed, this effect seems to be proportional to the charge applied, and is maximum when the PMT was previously kept in the dark, reaching significant gain variation peak amplitudes up to $\sim$ 8\%, asymptotically reduced to 3\%/2\% after 5/10 minutes respectively. It should be noted that such gain variations can be tracked in our system and, under on-sky observation conditions, this effect has never been observed with such an amplitude, and as it is only expected to affect significantly a small percentage of the observing time. We set a conservative overall limit over our measurements of $\sim$ 1\% for very bright targets (used to compute $V^2(0)_{i, j}$).
    \item PMT degradation. All PMTs undergo a slow process of gain degradation given the accumulated charge they are exposed to. We estimate, by assuming generous yearly charge exposures, a maximum yearly gain variation of 0.8\%. Calibration measurements will be able to correct for this effect, but as best calibrators are mainly observable in winter, a conservative 0.8\% systematic in the gain is assumed.
    \item Observations with different HV will lead to different gains. A gain calibration is performed as described in section \ref{sec:calibration} to account for small HV variations. With a constant photon flux, $log(DC)$ is expected to increase linearly with $log(HV)$. Very high current values are likely to escape this linear relation, and therefore would lead to different gains across our observations. To ensure this effect is not significant, calibration measurements over the HV and DC values we use for SII observations were performed, both to measure the relation between HV and DC for each pixel, and also ensure proportionality. No deviation from linearity was seen for any pixel over the safe current values employed.
\end{enumerate}

\subparagraph{Night-sky background DC subtraction:} As described in section \ref{sec:calibration}, the interferometry analysis needs to evaluate separately $DC(NSB)$ and $DC(Star)$. To do so, we simultaneously gather DC measurements from our signal and background pixels. Conversion factors are needed to calculate the equivalent DC(NSB) at the signal pixel. There are several systematic sources associated to this background subtraction: a) If stars of a comparable brightness than the NSB enter background pixels FoV, their evaluation of the background will be overestimated; b) Slowly evolving effects may also affect this conversion factors, such as dirt and dust on the surface of individual narrow-band filters (not affecting the optical bandwidth, but would affect the relative DC measured between different pixels). As most of our observations (the full reflector focusing light to a single pixel) allow us to have one signal and two background pixels simultaneously, we can directly evaluate with the gathered data what is the scale of the systematic we expect from a wrong background evaluation. It should be noted that this is expected to be very small for bright stars, but it will increase for fainter stars, in which DC(NSB) will be a significant fraction of DC(Star). The impact on the evaluation of the DC(star) is expected to be 1.5/3\% for bright/faint stars respectively (assuming faint stars are those fainter than $\sim$ 3.5 B mag). These are conservative values as during full-mirror observations we have 2 available background pixels: by selecting the one providing a lower DC(NSB), we suppress effect a), while by monitoring the ratio of DC(NSB) between background pixels allow us to update background-to-signal pixel conversion factors (and therefore, mitigate the effect b). 

\subparagraph{Residual electronic noise in the correlation:} The MAGIC counting house, where the correlator is located, is equipped with multiple electronic devices producing high-frequency noise, that could eventually affect the interferometry setup. Any residual noise in the correlation would add systematics to the visibility measurements by deviating Pearson's correlation from the true value. We have done extensive tests to estimate the scale of this residual correlation, which can be seen when no photons are being recorded (HV off, and therefore signals exclusively come from electronic noise from digitizers and correlators). We evaluate the scale of this systematic by integrating over long observing periods, and evaluate the resulting deviations with respect to 0 for $\rho(\tau)$ far from the region where we expect the signal. When integrating very long observing times ($>$ 50 h), a low-amplitude 1.67 MHz electronic noise is measured, stable in the $\rho(\tau) \cdot \sigma_1\cdot \sigma_2$ space (removing the normalization of Pearson's correlation). This systematic is strongly mitigated when adding a strong photon flux, and is therefore negligible for bright stars. This systematic, if not suppressed, would be significant when observing faint stars over dark conditions. By applying a 12 MHz cut in frequency (one order of magnitude narrower than our bandwidth) the residual electronic noise is strongly mitigated while correlation signals remain unaffected.

\begin{table}
\centering
\begin{tabular}{l|r} 
Systematic effect & Uncertainty \\
\hline
Electronic bandwidth & 0.5\% \\
Optical bandwidth & $<$ 1\% \\
Gain evolution of DC ADC branch & \\
- Seasonal temperature & Negligible \\
- Gain drift after DC jump & 1\% \\
- Long-term degradation & 0.8\% \\
- Deviations from linearity & Negligible \\
Residual electronic noise & Negligible \\
DC NSB substraction  & 1.5/3\% (B$_{mag}$ > 3.5) \\
\end{tabular}
\caption{Evaluated systematic uncertainties over squared visibility measurements identified to effect the MAGIC-SII system. \label{tab:systematics}}
\end{table}

\subparagraph{}The stability of the zero-baseline correlation was evaluated by testing different procedures to measure $V^2(0)_{i, j}$, each under different assumptions: i) Using eps CMa channel-pair wise data independently; ii) Using eps CMa data and assuming a common $\theta_{UD}$; iii) Using all available datasets from stars expected to be well described by a UD profile (removing fast rotators and spectroscopic binaries). In all cases, $V^2(0)_{i, j}$ are statistically consistent (largest variation of $\sim 2\%$). 

It should be noted that the scale in time in which these systematics take effect may be very different, and therefore it will affect the reconstructed stellar diameter measurements in different ways. Long-term PMT degradation, even if corrected with calibration measurements updating $V^2(0)_{i, j}$, will add a small systematic ($<$ 0.8\%) for summer sources (as most of the brightest reference stars are observable in winter). Effects expected to be transient over daily timescales (e.g. optical bandwidth deviations or some effects modifying PMT gains) are expected to cancel out when integrating datasets covering significantly longer time periods (e.g. when computing $V^2(0)_{i, j}$, or when measuring faint sources with observations spanning over multiple weeks/months). For the measurement of $V^2(0)_{i, j}$ we use observations  covering two different winters, each with multiple observing nights. In the case of eps CMa, $V^2(0)_{i, j}$ was measured simultaneously to $\theta$, reducing the chances of systematic differences between $V^2(0)$ and $V^2(d)$ measurements, and therefore the only sources capable of systematically deviating the measured stellar diameter are gain drift and DC(NSB) subtraction effects (a maximum combined uncertainty of 1.8\%). 

The expected systematic uncertainty for each measured $\theta$ were calculated by modifying squared visibility measurements in the most pessimistic way, and computing the limiting $\theta_{Syst, i}$ for each of these scenarios:
\begin{itemize}
    \item eps CMa: first/second half of visibility points were shifted up/down respectively by 1.8\%. 
    \item Others: $V^2(0)_{i, j}$ is shifted up/down by 1.8\% while the rest of $V^2$ measurements are shifted in the opposite direction by a brightness-dependent factor: 3.8\% if $B_{mag}$ > 3.5 and 2\% otherwise.
\end{itemize}

The reported systematic uncertainties in tables \ref{tab:calib_stars} and \ref{tab:candid_stars} refer to the difference between each $\theta_{Syst, i}$ and the best fit value, which is the largest deviation our systematics could cause in the most pessimistic scenario.

\subsection{Future prospects}
\label{sec:future}

We have presented an evaluation of the performance of the current interferometry setup in MAGIC and the first measurements of diameters of several stars. As shown in Fig. \ref{fig:RelativeErrorDiameters}, MAGIC can currently realistically target stars until $\sim$ 4 B mag, which means that we cannot compete yet with other long-baseline optical interferometers such as CHARA \citep{CHARA_VEGA}. But these results prove the potential of planned improvements to boost sensitivity and angular resolution, described in more detail in \cite{spie2022}.

\subparagraph{Bandwidth improvements:} As discussed in section \ref{sec:photon_transmission}, the current system bandwidth is limited by the digitizer used. We expect a factor $\sim \sqrt{2}$ improvement in sensitivity by upgrading digitizers to the next generation (Spectrum M5i.3321-x16). Improvements beyond this point would require faster photo-detectors and upgraded signal transmission. An alternative approach has also been tested during the last years: a faster readout based on a commercial 4 GHz sampling rate ADC coupled to an FPGA (Xilinx ZYNQ UltraScale+ RFSoC ZU28DR) which performs the correlation real-time. This readout currently operates in parallel to our nominal readout. Significant correlation signals have been detected although the system suffers from a strong correlated noise, with prominent components at frequencies beyond the bandpass of our nominal readout.

\subparagraph{Photo-detection efficiency:} a relatively straightforward way of improving the current system sensitivity would be to upgrade MAGIC photo-detectors to a newer generation of these detectors, as those used within the LST-1 camera. If such an array was implemented, the QE would increase from the current $\sim$ 32\% to 40\%.

\begin{figure}
\centering
\includegraphics[width=\columnwidth]{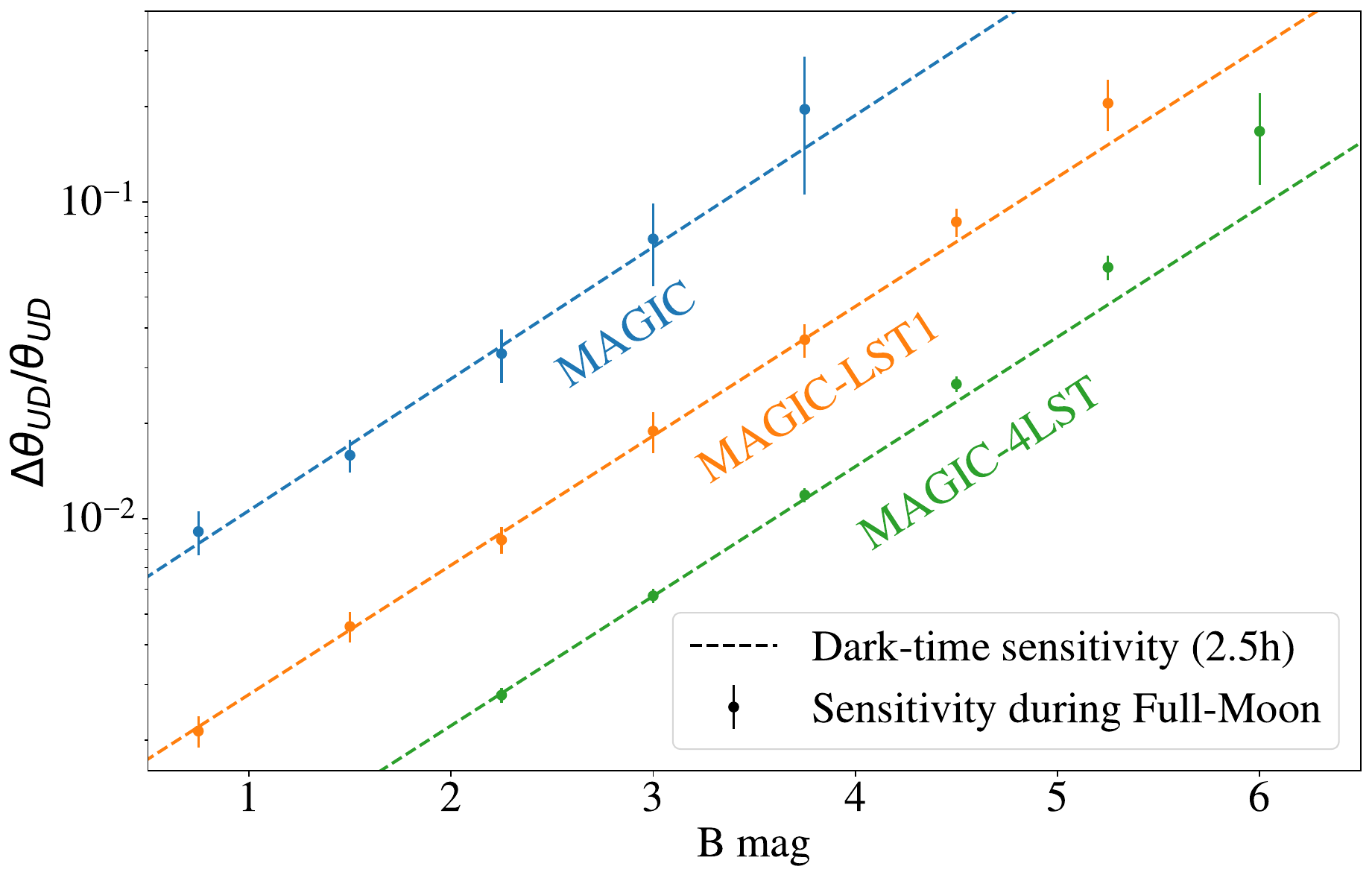}
\caption{Relative uncertainty of measured stellar diameters as a function of their B magnitude for the current MAGIC interferometer, an extension to MAGIC and LST-1, and a further extension to MAGIC and the four LSTs in La Palma. Dashed lines indicate sensitivity during dark time, while solid points estimate the sensitivity under conservative very-high NSB illumination levels (both for a 2.5 h observing time). The simulated stellar position and diameter is the one from gam CrV. \label{fig:error_diameter_MAGIC_4LST}}
\end{figure}

\subparagraph{Additional telescopes:} The first 23 m large-sized telescope prototype (dubbed LST-1) of the future Cherenkov Telescope Array Observatory (CTAO) was inaugurated in 2018 at a distance of $\sim$100 m from the MAGIC telescopes \citep{lst}. Three more telescopes with almost identical characteristics (LST2-4) are under construction and assembly, scheduled to become operational in 2025. In addition to mirror area (370 m$^2$), photo-detectors QE ($\sim$ 40\%), and total optical efficiency ($\sim$ 60\%) are expected to be significantly better than the one from MAGIC. This increased photon-detection efficiency would dramatically improve sensitivity and the simultaneous UV coverage of the array. The expected relative errors as a function of stellar magnitude of MAGIC-LST1 and MAGIC-LST1-4 arrays is shown in Fig. \ref{fig:error_diameter_MAGIC_4LST} \citep{spie2022}. Minimal modifications in one cluster of LST-1 camera electronics have been designed and tested to enable optical transmission of a single pixel data to the MAGIC correlator. Adding additional telescopes to the array imply adding more input channels to the current GPU-based correlator. As described in section \ref{sec:digitizer_correlator}, the current system is already able to handle four input channels, which means no upgrades are necessary for the MAGIC-LST-1 array. However we typically operate with two pixels in all telescopes, to allow sub-reflector and chess-board observations. For this reason, the optimal implementation of adding the four LSTs would require a total of 12 input channels. A concept for a new GPU-based scalable correlator is currently being tested, described in detail by \cite{spie2022}. If successful, this concept with be capable of handling all input signals coming from the full Northern CTAO.

\section{Conclusions and discussion}

In this publication, we have demonstrated the feasibility of using IACTs as optical intensity interferometry arrays with only minor hardware additions. Given the narrow optical bandpass employed for these observations, they may be performed during full-Moon periods, times in which IACTs are generally not able to perform VHE gamma-ray astronomy. This breakthrough expands the scientific impact of IACT facilities and also increases their operational duty cycle.

We have found that the measurements made by the MAGIC-SII system are in good agreement with the measurements obtained by other observatories (see Table \ref{tab:calib_stars}). Having demonstrated the capabilities of the system, we were thus able to reliably measure the diameters of 13 stars in the 400-440 nm band for the first time (see Table \ref{tab:candid_stars}). Furthermore, we have shown that the sensitivity of the system is already capable of reaching relative errors at the few percent level over reasonable  observing times (e.g. eps CMa or eta UMa). This level of precision, in principle, already allows for the study of the oblateness of fast-rotating stars. Measuring the oblateness of particular stars (such as blue supergiants, hypergiants, and Wolf-Rayet stars) allows to constrain their rotational speed, a crucial parameter that affects their structure and evolution. 
In order to allow for this or any other non-radially symmetric model fitting procedure, the MAGIC collaboration will release future MAGIC-SII data products using optical interferometry standards, such as OIFITS \citep{oifits}.

In the near future, we will also be able to perform the first test observations by including the LST-1 in the array. The large increase in the expected S/N will realistically allow us to increase the precision of the diameter measurements, enabling us to achieve relative statistical errors below 1\% over a large sample of stars. At that point, we expect to reach the scale of the systematics affecting our current analysis. The study of these systematics will also be easier by having a better S/N: shorter observation times will be required to make measurements of the correlation and, given the multiple simultaneous baselines and orientations covered, more stars can be employed to constrain them.

In conclusion, the results of the observations shown in this paper are consistent with expectations. Although they may not yet be competitive with other long-baseline interferometers (e.g., CHARA), this work demonstrates that the MAGIC-SII system will certainly be competitive if other telescopes belonging to the future Cherenkov Telescope Array Observatory northern array are included, as shown in Fig. \ref{fig:error_diameter_MAGIC_4LST}. Adding additional telescopes to the array using the intensity interferometry technique is technically simpler than in the case of classical optical interferometry, which means it may become a realistic avenue for performing optical interferometry with inter-telescope distances well beyond the km in the coming years. Although sensitivity may not extend beyond the 7-8 magnitude scale for such simple implementations (given the poor optical Point Spread Function of IACTs), the UV coverage achieved by large and dense arrays of IACTs would be unprecedented in the optical regime.

\section*{Acknowledgements}
We would like to thank the Instituto de Astrof\'{\i}sica de Canarias for the excellent working conditions at the Observatorio del Roque de los Muchachos in La Palma. The financial support of the German BMBF, MPG and HGF; the Italian INFN and INAF; the Swiss National Fund SNF; the grants PID2019-104114RB-C31, PID2019-104114RB-C32, PID2019-104114RB-C33, PID2019-105510GB-C31, PID2019-107847RB-C41, PID2019-107847RB-C42, PID2019-107847RB-C44, PID2019-107988GB-C22, PID2022-136828NB-C41, PID2022-137810NB-C22, PID2022-138172NB-C41, PID2022-138172NB-C42, PID2022-138172NB-C43, PID2022-139117NB-C41, PID2022-139117NB-C42, PID2022-139117NB-C43, PID2022-139117NB-C44 funded by the Spanish MCIN/AEI/ 10.13039/501100011033 and “ERDF A way of making Europe”; the Indian Department of Atomic Energy; the Japanese ICRR, the University of Tokyo, JSPS, and MEXT; the Bulgarian Ministry of Education and Science, National RI Roadmap Project DO1-400/18.12.2020 and the Academy of Finland grant nr. 320045 is gratefully acknowledged. This work was also been supported by Centros de Excelencia ``Severo Ochoa'' y Unidades ``Mar\'{\i}a de Maeztu'' program of the Spanish MCIN/AEI/ 10.13039/501100011033 (CEX2019-000920-S, CEX2019-000918-M, CEX2021-001131-S) and by the CERCA institution and grants 2021SGR00426 and 2021SGR00773 of the Generalitat de Catalunya; by the Croatian Science Foundation (HrZZ) Project IP-2022-10-4595 and the University of Rijeka Project uniri-prirod-18-48; by the Deutsche Forschungsgemeinschaft (SFB1491) and by the Lamarr-Institute for Machine Learning and Artificial Intelligence; by the Polish Ministry Of Education and Science grant No. 2021/WK/08; and by the Brazilian MCTIC, CNPq and FAPERJ.
Funded/Co-funded by the European Union (ERC, MicroStars, 101076533). Views and opinions expressed are however those of the author(s) only and do not necessarily reflect those of the European Union or the European Research Council. Neither the European Union nor the granting authority can be held responsible for them.

\section*{Author contributions}
This work is the culmination of a collaborative effort by the MAGIC Interferometry team, consisting of the following dedicated members: V. A. Acciari, G. Chon, A. Cifuentes, E. Colombo, J. Cortina, C. Delgado, D. Fink, M. Fiori, T. Hassan, I. Jiménez Martínez, E. Lyard, S. Mangano, R. Mirzoyan, T. Njoh, N. Produit, J. J. Rodriguez-Vazquez, P. Saha, T. Schweizer, R. Walter and L. Zampieri.
C. Díaz and M. Polo designed, produced and installed the filter holder, and M. Lobo measured the transmission of the optical filters. C. Delgado designed the concept of the correlator. J. J. Rodríguez is the main developer and maintainer of the adquisition software. V. A. Acciari, E. Colombo and T. Schweizer configured the correlator and performed integration tests on site. D. Fink implemented the signal acquisition and performed
laboratory measurements to evaluate the gain evolution. I. Jiménez-Martínez and A. Cifuentes were in charge of target selection and scheduling observations. T. Hassan developed the on-site analysis. J. Cortina and T. Hassan developed the communications and integration with MAGIC central control. T. Hassan, I. Jiménez-Martínez, M. Fiori and C. Wunderlich developed independent high-level analysis libraries to perform data calibration and stellar diameter measurements. The rest of the authors have contributed in one or several of the following ways: design, construction, maintenance and operation of the instrument(s) used to acquire the data; preparation and/ or evaluation of the observation proposals; data acquisition, processing, calibration and/ or reduction; production of analysis tools and/ or related Monte Carlo simulations; overall discussions about the contents of the draft, as well as related refinements in the descriptions.

\section*{Data Availability}

Access to MAGIC data is reserved to members of the MAGIC Collaboration.  The MAGIC-SII team is working towards producing high-level data products following optical interferometry standards (OIFITS). These high-level data products may be shared upon request.



\bibliographystyle{mnras}
\bibliography{bibliography} 

\begin{thebibliography}{}
\makeatletter
\relax
\def\mn@urlcharsother{\let\do\@makeother \do\$\do\&\do\#\do\^\do\_\do\%\do\~}
\def\mn@doi{\begingroup\mn@urlcharsother \@ifnextchar [ {\mn@doi@}
  {\mn@doi@[]}}
\def\mn@doi@[#1]#2{\def\@tempa{#1}\ifx\@tempa\@empty \href
  {http://dx.doi.org/#2} {doi:#2}\else \href {http://dx.doi.org/#2} {#1}\fi
  \endgroup}
\def\mn@eprint#1#2{\mn@eprint@#1:#2::\@nil}
\def\mn@eprint@arXiv#1{\href {http://arxiv.org/abs/#1} {{\tt arXiv:#1}}}
\def\mn@eprint@dblp#1{\href {http://dblp.uni-trier.de/rec/bibtex/#1.xml}
  {dblp:#1}}
\def\mn@eprint@#1:#2:#3:#4\@nil{\def\@tempa {#1}\def\@tempb {#2}\def\@tempc
  {#3}\ifx \@tempc \@empty \let \@tempc \@tempb \let \@tempb \@tempa \fi \ifx
  \@tempb \@empty \def\@tempb {arXiv}\fi \@ifundefined
  {mn@eprint@\@tempb}{\@tempb:\@tempc}{\expandafter \expandafter \csname
  mn@eprint@\@tempb\endcsname \expandafter{\@tempc}}}

\bibitem[\protect\citeauthoryear{Abeysekara \& al.}{Abeysekara \&
  al.}{2020}]{VERITAS}
Abeysekara A.~U.,  al. 2020, Nature Astronomy, 4, 1164

\bibitem[\protect\citeauthoryear{Acciari \& al.}{Acciari \&
  al.}{2020a}]{spie2020}
Acciari V.~A.,  al. 2020a, in Optical and Infrared Interferometry and Imaging
  VII.

\bibitem[\protect\citeauthoryear{Acciari \& al.}{Acciari \&
  al.}{2020b}]{magic_2019}
Acciari V.~A.,  al. 2020b, MNRAS, 491-2, 1540–1547

\bibitem[\protect\citeauthoryear{{Aleksi{\'c}} et~al.,}{{Aleksi{\'c}}
  et~al.}{2012}]{performance_paper}
{Aleksi{\'c}} J.,  et~al., 2012, \mn@doi [Astroparticle Physics]
  {10.1016/j.astropartphys.2011.11.007}, \href
  {https://ui.adsabs.harvard.edu/abs/2012APh....35..435A} {35, 435}

\bibitem[\protect\citeauthoryear{{Aleksi{\'c}} et~al.,}{{Aleksi{\'c}}
  et~al.}{2016a}]{Aleksic2016}
{Aleksi{\'c}} J.,  et~al., 2016a, \mn@doi [Astroparticle Physics]
  {10.1016/j.astropartphys.2015.04.004}, \href
  {https://ui.adsabs.harvard.edu/abs/2016APh....72...61A} {72, 61}

\bibitem[\protect\citeauthoryear{{Aleksi{\'c}} et~al.,}{{Aleksi{\'c}}
  et~al.}{2016b}]{performance_upgrade}
{Aleksi{\'c}} J.,  et~al., 2016b, \mn@doi [Astroparticle Physics]
  {10.1016/j.astropartphys.2015.02.005}, \href
  {https://ui.adsabs.harvard.edu/abs/2016APh....72...76A} {72, 76}

\bibitem[\protect\citeauthoryear{{Allende Prieto} \& {Lambert}}{{Allende
  Prieto} \& {Lambert}}{1999}]{physical_pars_4}
{Allende Prieto} C.,  {Lambert} D.~L.,  1999, \mn@doi [\aap]
  {10.48550/arXiv.astro-ph/9911002}, \href
  {https://ui.adsabs.harvard.edu/abs/1999A&A...352..555A} {352, 555}

\bibitem[\protect\citeauthoryear{{Anderson} \& {Francis}}{{Anderson} \&
  {Francis}}{2012}]{physical_pars_1}
{Anderson} E.,  {Francis} C.,  2012, \mn@doi [Astronomy Letters]
  {10.1134/S1063773712050015}, \href
  {https://ui.adsabs.harvard.edu/abs/2012AstL...38..331A} {38, 331}

\bibitem[\protect\citeauthoryear{{Baines}, {Armstrong}, {Schmitt}, {Zavala},
  {Benson}, {Hutter}, {Tycner}  \& {van Belle}}{{Baines}
  et~al.}{2018}]{physical_pars_2}
{Baines} E.~K.,  {Armstrong} J.~T.,  {Schmitt} H.~R.,  {Zavala} R.~T.,
  {Benson} J.~A.,  {Hutter} D.~J.,  {Tycner} C.,   {van Belle} G.~T.,  2018,
  \mn@doi [\aj] {10.3847/1538-3881/aa9d8b}, \href
  {https://ui.adsabs.harvard.edu/abs/2018AJ....155...30B} {155, 30}

\bibitem[\protect\citeauthoryear{{Bastieri} et~al.,}{{Bastieri}
  et~al.}{2008}]{reflecting_surface}
{Bastieri} D.,  et~al., 2008, in International Cosmic Ray Conference. pp
  1547--1550 (\mn@eprint {arXiv} {0709.1372}),
  \mn@doi{10.48550/arXiv.0709.1372}

\bibitem[\protect\citeauthoryear{{Baumgartner} et~al.,}{{Baumgartner}
  et~al.}{2020}]{2020MNRAS.498.4577B}
{Baumgartner} S.,  et~al., 2020, \mn@doi [\mnras] {10.1093/mnras/staa2638},
  \href {https://ui.adsabs.harvard.edu/abs/2020MNRAS.498.4577B} {498, 4577}

\bibitem[\protect\citeauthoryear{{Baym}}{{Baym}}{1998}]{1998AcPPB..29.1839B}
{Baym} G.,  1998, \mn@doi [Acta Physica Polonica B]
  {10.48550/arXiv.nucl-th/9804026}, \href
  {https://ui.adsabs.harvard.edu/abs/1998AcPPB..29.1839B} {29, 1839}

\bibitem[\protect\citeauthoryear{{Berger} \& {Segransan}}{{Berger} \&
  {Segransan}}{2007}]{visibility_modeling}
{Berger} J.~P.,  {Segransan} D.,  2007, \mn@doi [\nar]
  {10.1016/j.newar.2007.06.003}, \href
  {https://ui.adsabs.harvard.edu/abs/2007NewAR..51..576B} {51, 576}

\bibitem[\protect\citeauthoryear{{Biland}, {Garczarczyk}, {Anderhub},
  {Danielyan}, {Hakobyan}, {Lorenz}  \& {Mirzoyan}}{{Biland}
  et~al.}{2008}]{amc_icrc}
{Biland} A.,  {Garczarczyk} M.,  {Anderhub} H.,  {Danielyan} V.,  {Hakobyan}
  D.,  {Lorenz} E.,   {Mirzoyan} R.,  2008, in International Cosmic Ray
  Conference. pp 1353--1356 (\mn@eprint {arXiv} {0709.1574}),
  \mn@doi{10.48550/arXiv.0709.1574}

\bibitem[\protect\citeauthoryear{Bitossi, Cecchi, Paoletti, Pegna, Turini,
  Barcelo  \& Illa}{Bitossi et~al.}{2007}]{MII_daq}
Bitossi M.,  Cecchi R.,  Paoletti R.,  Pegna R.,  Turini N.,  Barcelo M.,
  Illa J.~M.,  2007, in 2007 4th IEEE Workshop on Intelligent Data Acquisition
  and Advanced Computing Systems: Technology and Applications. pp 60--62,
  \mn@doi{10.1109/IDAACS.2007.4488373}

\bibitem[\protect\citeauthoryear{Bonneau et~al.,}{Bonneau
  et~al.}{2006}]{bonneau2006searchcal}
Bonneau D.,  et~al., 2006, Astronomy \& Astrophysics, 456, 789

\bibitem[\protect\citeauthoryear{Bonneau, Delfosse, Mourard, Lafrasse, Mella,
  Cetre, Clausse  \& Zins}{Bonneau et~al.}{2011}]{bonneau2011searchcal}
Bonneau D.,  Delfosse X.,  Mourard D.,  Lafrasse S.,  Mella G.,  Cetre S.,
  Clausse J.-M.,   Zins G.,  2011, Astronomy \& Astrophysics, 535, A53

\bibitem[\protect\citeauthoryear{{Bourdarot}, {Guillet de Chatellus}  \&
  {Berger}}{{Bourdarot} et~al.}{2020}]{2020A&A...639A..53B}
{Bourdarot} G.,  {Guillet de Chatellus} H.,   {Berger} J.~P.,  2020, \mn@doi
  [\aap] {10.1051/0004-6361/201937368}, \href
  {https://ui.adsabs.harvard.edu/abs/2020A&A...639A..53B} {639, A53}

\bibitem[\protect\citeauthoryear{Bourges, Mella, Lafrasse, Duvert, Chelli,
  Le~Bouquin, Delfosse  \& Chesneau}{Bourges et~al.}{2017}]{bourges2017vizier}
Bourges L.,  Mella G.,  Lafrasse S.,  Duvert G.,  Chelli A.,  Le~Bouquin J.-B.,
   Delfosse X.,   Chesneau O.,  2017, VizieR Online Data Catalog, pp II--346

\bibitem[\protect\citeauthoryear{{Cardiel} et~al.,}{{Cardiel}
  et~al.}{2021}]{physical_pars_5}
{Cardiel} N.,  et~al., 2021, \mn@doi [\mnras] {10.1093/mnras/stab997}, \href
  {https://ui.adsabs.harvard.edu/abs/2021MNRAS.504.3730C} {504, 3730}

\bibitem[\protect\citeauthoryear{{Cavazzani}, {Ortolani}  \&
  {Barbieri}}{{Cavazzani} et~al.}{2012}]{atmosphere_times}
{Cavazzani} S.,  {Ortolani} S.,   {Barbieri} C.,  2012, \mn@doi [\mnras]
  {10.1111/j.1365-2966.2011.19883.x}, \href
  {https://ui.adsabs.harvard.edu/abs/2012MNRAS.419.2349C} {419, 2349}

\bibitem[\protect\citeauthoryear{{Challouf} et~al.,}{{Challouf}
  et~al.}{2014}]{chara_gam_lyr}
{Challouf} M.,  et~al., 2014, \mn@doi [\aap] {10.1051/0004-6361/201423772},
  \href {https://ui.adsabs.harvard.edu/abs/2014A&A...570A.104C} {570, A104}

\bibitem[\protect\citeauthoryear{{Che} et~al.,}{{Che}
  et~al.}{2011}]{2011ApJ...732...68C}
{Che} X.,  et~al., 2011, \mn@doi [\apj] {10.1088/0004-637X/732/2/68}, \href
  {https://ui.adsabs.harvard.edu/abs/2011ApJ...732...68C} {732, 68}

\bibitem[\protect\citeauthoryear{{Claret} \& {Bloemen}}{{Claret} \&
  {Bloemen}}{2011}]{ld_linear_coeffs}
{Claret} A.,  {Bloemen} S.,  2011, \mn@doi [\aap]
  {10.1051/0004-6361/201116451}, \href
  {https://ui.adsabs.harvard.edu/abs/2011A&A...529A..75C} {529, A75}

\bibitem[\protect\citeauthoryear{{Cortina} et~al.,}{{Cortina}
  et~al.}{2022}]{spie2022}
{Cortina} J.,  et~al., 2022, in {M{\'e}rand} A.,  {Sallum} S.,
  {Sanchez-Bermudez} J.,  eds,  Society of Photo-Optical Instrumentation
  Engineers (SPIE) Conference Series Vol. 12183, Optical and Infrared
  Interferometry and Imaging VIII. p. 121830C (\mn@eprint {arXiv}
  {2209.14844}), \mn@doi{10.1117/12.2629876}

\bibitem[\protect\citeauthoryear{{Dazzi} et~al.,}{{Dazzi}
  et~al.}{2021}]{trigger}
{Dazzi} F.,  et~al., 2021, \mn@doi [IEEE Transactions on Nuclear Science]
  {10.1109/TNS.2021.3079262}, \href
  {https://ui.adsabs.harvard.edu/abs/2021ITNS...68.1473D} {68, 1473}

\bibitem[\protect\citeauthoryear{Delgado \& al.}{Delgado \&
  al.}{2021}]{icrc2021}
Delgado C.,  al. 2021, in Proc. 37th International Cosmic Ray Conference,
  Berlin 2021.

\bibitem[\protect\citeauthoryear{{Dravins}, {LeBohec}, {Jensen}  \&
  {Nu\~nez}}{{Dravins} et~al.}{2012}]{Dravins2012}
{Dravins} D.,  {LeBohec} S.,  {Jensen} S.,   {Nu\~nez} P.~D.,  2012, New Astr.
  Rev., 56, 143

\bibitem[\protect\citeauthoryear{{Dravins}, {LeBohec}, {Jensen}, {Nu{\~n}ez}
  \& {CTA Consortium}}{{Dravins} et~al.}{2013}]{Dravins2013}
{Dravins} D.,  {LeBohec} S.,  {Jensen} H.,  {Nu{\~n}ez} P.~D.,   {CTA
  Consortium} 2013, \mn@doi [Astroparticle Physics]
  {10.1016/j.astropartphys.2012.04.017}, \href
  {https://ui.adsabs.harvard.edu/abs/2013APh....43..331D} {43, 331}

\bibitem[\protect\citeauthoryear{{Duvert}, {Young}  \& {Hummel}}{{Duvert}
  et~al.}{2017}]{oifits}
{Duvert} G.,  {Young} J.,   {Hummel} C.~A.,  2017, \mn@doi [\aap]
  {10.1051/0004-6361/201526405}, \href
  {https://ui.adsabs.harvard.edu/abs/2017A&A...597A...8D} {597, A8}

\bibitem[\protect\citeauthoryear{{Eisenhauer}, {Monnier}  \&
  {Pfuhl}}{{Eisenhauer} et~al.}{2023}]{Eisenhauer_review}
{Eisenhauer} F.,  {Monnier} J.~D.,   {Pfuhl} O.,  2023, \mn@doi [\araa]
  {10.1146/annurev-astro-121622-045019}, \href
  {https://ui.adsabs.harvard.edu/abs/2023ARA&A..61..237E} {61, 237}

\bibitem[\protect\citeauthoryear{{Event Horizon Telescope Collaboration:
  Akiyama} et~al.,}{{Event Horizon Telescope Collaboration: Akiyama}
  et~al.}{2022}]{2022ApJ...930L..13E}
{Event Horizon Telescope Collaboration: Akiyama} K.,  et~al., 2022, \mn@doi
  [\apjl] {10.3847/2041-8213/ac6675}, \href
  {https://ui.adsabs.harvard.edu/abs/2022ApJ...930L..13E} {930, L13}

\bibitem[\protect\citeauthoryear{{Fiori}, {Naletto}, {Zampieri},
  {Mart{\'\i}nez}  \& {Wunderlich}}{{Fiori} et~al.}{2022}]{fiori}
{Fiori} M.,  {Naletto} G.,  {Zampieri} L.,  {Mart{\'\i}nez} I.~J.,
  {Wunderlich} C.,  2022, \mn@doi [\aap] {10.1051/0004-6361/202244094}, \href
  {https://ui.adsabs.harvard.edu/abs/2022A&A...666A..48F} {666, A48}

\bibitem[\protect\citeauthoryear{{GRAVITY Collaboration: Abuter}
  et~al.,}{{GRAVITY Collaboration: Abuter} et~al.}{2022}]{2022A&A...657A..82G}
{GRAVITY Collaboration: Abuter} R.,  et~al., 2022, \mn@doi [\aap]
  {10.1051/0004-6361/202142459}, \href
  {https://ui.adsabs.harvard.edu/abs/2022A&A...657A..82G} {657, A82}

\bibitem[\protect\citeauthoryear{{Gordon}, {Gies}, {Schaefer}, {Huber}  \&
  {Ireland}}{{Gordon} et~al.}{2019}]{chara_eta_uma}
{Gordon} K.~D.,  {Gies} D.~R.,  {Schaefer} G.~H.,  {Huber} D.,   {Ireland} M.,
  2019, \mn@doi [\apj] {10.3847/1538-4357/ab04b2}, \href
  {https://ui.adsabs.harvard.edu/abs/2019ApJ...873...91G} {873, 91}

\bibitem[\protect\citeauthoryear{{Gori} et~al.,}{{Gori}
  et~al.}{2021}]{i3t_gori2021}
{Gori} P.-M.,  et~al., 2021, \mn@doi [\mnras] {10.1093/mnras/stab1424}, \href
  {https://ui.adsabs.harvard.edu/abs/2021MNRAS.505.2328G} {505, 2328}

\bibitem[\protect\citeauthoryear{{Hanbury Brown}}{{Hanbury
  Brown}}{1974}]{HBbook1974}
{Hanbury Brown} R.,  1974, The Intensity Interferometer: Its Application to
  Astronomy.
Taylor \& Francis, London

\bibitem[\protect\citeauthoryear{{Hanbury Brown} \& {Twiss}}{{Hanbury Brown} \&
  {Twiss}}{1958}]{HBT1958}
{Hanbury Brown} R.,  {Twiss} R.~Q.,  1958, \mn@doi [Proceedings of the Royal
  Society of London Series A] {10.1098/rspa.1958.0240}, \href
  {https://ui.adsabs.harvard.edu/abs/1958RSPSA.248..222B} {248, 222}

\bibitem[\protect\citeauthoryear{{Hanbury Brown}, {Davis}, {Lake}  \&
  {Thompson}}{{Hanbury Brown} et~al.}{1974a}]{HB_LD_model}
{Hanbury Brown} R.,  {Davis} J.,  {Lake} R.~J.~W.,   {Thompson} R.~J.,  1974a,
  \mn@doi [\mnras] {10.1093/mnras/167.3.475}, \href
  {https://ui.adsabs.harvard.edu/abs/1974MNRAS.167..475H} {167, 475}

\bibitem[\protect\citeauthoryear{{Hanbury Brown}, {Davis}  \& {Allen}}{{Hanbury
  Brown} et~al.}{1974b}]{HB1974}
{Hanbury Brown} R.,  {Davis} J.,   {Allen} L.~R.,  1974b, MNRAS, 167-1, 121

\bibitem[\protect\citeauthoryear{{Horch}, {Weiss}, {Klaucke}, {Pellegrino}  \&
  {Rupert}}{{Horch} et~al.}{2022}]{SCSI}
{Horch} E.~P.,  {Weiss} S.~A.,  {Klaucke} P.~M.,  {Pellegrino} R.~A.,
  {Rupert} J.~D.,  2022, \mn@doi [\aj] {10.3847/1538-3881/ac43bb}, \href
  {https://ui.adsabs.harvard.edu/abs/2022AJ....163...92H} {163, 92}

\bibitem[\protect\citeauthoryear{{Karl} et~al.,}{{Karl} et~al.}{2022}]{HESS1}
{Karl} S.,  et~al., 2022, \mn@doi [\mnras] {10.1093/mnras/stac489}, \href
  {https://ui.adsabs.harvard.edu/abs/2022MNRAS.512.1722K} {512, 1722}

\bibitem[\protect\citeauthoryear{{Kieda}, {Davis}, {LeBohec}, {Lisa}  \& {et
  al.}}{{Kieda} et~al.}{2022}]{VERITAS3}
{Kieda} D.,  {Davis} J.,  {LeBohec} T.,  {Lisa} M.,   {et al.} 2022, in 37th
  International Cosmic Ray Conference. p.~803 (\mn@eprint {arXiv}
  {2108.09774}), \mn@doi{10.22323/1.395.0803}

\bibitem[\protect\citeauthoryear{{Labeyrie}}{{Labeyrie}}{2021}]{2021RSPTA.37990570L}
{Labeyrie} A.,  2021, \mn@doi [Philosophical Transactions of the Royal Society
  of London Series A] {10.1098/rsta.2019.0570}, \href
  {https://ui.adsabs.harvard.edu/abs/2021RSPTA.37990570L} {379, 20190570}

\bibitem[\protect\citeauthoryear{{Le Bohec} \& {Holder}}{{Le Bohec} \&
  {Holder}}{2006}]{2006ApJ...649..399L}
{Le Bohec} S.,  {Holder} J.,  2006, \mn@doi [\apj] {10.1086/506379}, \href
  {https://ui.adsabs.harvard.edu/abs/2006ApJ...649..399L} {649, 399}

\bibitem[\protect\citeauthoryear{{Matthews} et~al.,}{{Matthews}
  et~al.}{2023}]{Nice}
{Matthews} N.,  et~al., 2023, \mn@doi [\aj] {10.3847/1538-3881/acb142}, \href
  {https://ui.adsabs.harvard.edu/abs/2023AJ....165..117M} {165, 117}

\bibitem[\protect\citeauthoryear{Mazin \& al.}{Mazin \& al.}{2021}]{lst}
Mazin D.,  al. 2021, in Proc. 37th International Cosmic Ray Conference, Berlin
  2021.

\bibitem[\protect\citeauthoryear{{Michelson} \& {Pease}}{{Michelson} \&
  {Pease}}{1921}]{1921ApJ....53..249M}
{Michelson} A.~A.,  {Pease} F.~G.,  1921, \mn@doi [\apj] {10.1086/142603},
  \href {https://ui.adsabs.harvard.edu/abs/1921ApJ....53..249M} {53, 249}

\bibitem[\protect\citeauthoryear{{Montarg{\`e}s} et~al.,}{{Montarg{\`e}s}
  et~al.}{2021}]{2021sf2a.conf...13M}
{Montarg{\`e}s} M.,  et~al., 2021, in {Siebert} A.,  et~al., eds, SF2A-2021:
  Proceedings of the Annual meeting of the French Society of Astronomy and
  Astrophysics. Eds.: A. Siebert. pp 13--18

\bibitem[\protect\citeauthoryear{{Mourard} et~al.,}{{Mourard}
  et~al.}{2009}]{CHARA_VEGA}
{Mourard} D.,  et~al., 2009, \mn@doi [\aap] {10.1051/0004-6361/200913016},
  \href {https://ui.adsabs.harvard.edu/abs/2009A&A...508.1073M} {508, 1073}

\bibitem[\protect\citeauthoryear{{Nardetto} et~al.,}{{Nardetto}
  et~al.}{2016}]{2016A&A...593A..45N}
{Nardetto} N.,  et~al., 2016, \mn@doi [\aap] {10.1051/0004-6361/201528005},
  \href {https://ui.adsabs.harvard.edu/abs/2016A&A...593A..45N} {593, A45}

\bibitem[\protect\citeauthoryear{{Newville}, {Stensitzki}, {Allen}, {Rawlik},
  {Ingargiola}  \& {Nelson}}{{Newville} et~al.}{2016}]{lmfit}
{Newville} M.,  {Stensitzki} T.,  {Allen} D.~B.,  {Rawlik} M.,  {Ingargiola}
  A.,   {Nelson} A.,  2016, {Lmfit: Non-Linear Least-Square Minimization and
  Curve-Fitting for Python}, Astrophysics Source Code Library, record
  ascl:1606.014 (\mn@eprint {ascl} {1606.014})

\bibitem[\protect\citeauthoryear{{Nu{\~n}ez} \& {Domiciano de
  Souza}}{{Nu{\~n}ez} \& {Domiciano de Souza}}{2015}]{2015MNRAS.453.1999N}
{Nu{\~n}ez} P.~D.,  {Domiciano de Souza} A.,  2015, \mn@doi [\mnras]
  {10.1093/mnras/stv1719}, \href
  {https://ui.adsabs.harvard.edu/abs/2015MNRAS.453.1999N} {453, 1999}

\bibitem[\protect\citeauthoryear{{Soubiran}, {Le Campion}, {Brouillet}  \&
  {Chemin}}{{Soubiran} et~al.}{2016}]{physical_pars_3}
{Soubiran} C.,  {Le Campion} J.-F.,  {Brouillet} N.,   {Chemin} L.,  2016,
  \mn@doi [\aap] {10.1051/0004-6361/201628497}, \href
  {https://ui.adsabs.harvard.edu/abs/2016A&A...591A.118S} {591, A118}

\bibitem[\protect\citeauthoryear{{Swihart}, {Garcia}, {Stassun}, {van Belle},
  {Mutterspaugh}  \& {Elias}}{{Swihart} et~al.}{2017}]{2017AJ....153...16S}
{Swihart} S.~J.,  {Garcia} E.~V.,  {Stassun} K.~G.,  {van Belle} G.,
  {Mutterspaugh} M.~W.,   {Elias} N.,  2017, \mn@doi [\aj]
  {10.3847/1538-3881/153/1/16}, \href
  {https://ui.adsabs.harvard.edu/abs/2017AJ....153...16S} {153, 16}

\bibitem[\protect\citeauthoryear{{Wallace}}{{Wallace}}{1994}]{TPOINT}
{Wallace} P.~T.,  1994, Starlink User Note, \href
  {https://ui.adsabs.harvard.edu/abs/1994StaUN.100.....W} {100}

\makeatother
\end{thebibliography}

\section*{Affiliations}
$^{1}$ {Japanese MAGIC Group: Institute for Cosmic Ray Research (ICRR), The University of Tokyo, Kashiwa, 277-8582 Chiba, Japan} \\
$^{2}$ {ETH Z\"urich, CH-8093 Z\"urich, Switzerland} \\
$^{3}$ {Instituto de Astrof\'isica de Canarias and Dpto. de  Astrof\'isica, Universidad de La Laguna, E-38200, La Laguna, Tenerife, Spain} \\
$^{4}$ {Universitat de Barcelona, ICCUB, IEEC-UB, E-08028 Barcelona, Spain} \\
$^{5}$ {Instituto de Astrof\'isica de Andaluc\'ia-CSIC, Glorieta de la Astronom\'ia s/n, 18008, Granada, Spain} \\
$^{6}$ {National Institute for Astrophysics (INAF), I-00136 Rome, Italy} \\
$^{7}$ {Universit\`a di Udine and INFN Trieste, I-33100 Udine, Italy} \\
$^{8}$ {Max-Planck-Institut f\"ur Physik, D-85748 Garching, Germany} \\
$^{9}$ {Universit\`a di Padova and INFN, I-35131 Padova, Italy} \\
$^{10}$ {Institut de F\'isica d'Altes Energies (IFAE), The Barcelona Institute of Science and Technology (BIST), E-08193 Bellaterra (Barcelona), Spain} \\
$^{11}$ {Croatian MAGIC Group: University of Zagreb, Faculty of Electrical Engineering and Computing (FER), 10000 Zagreb, Croatia} \\
$^{12}$ {IPARCOS Institute and EMFTEL Department, Universidad Complutense de Madrid, E-28040 Madrid, Spain} \\
$^{13}$ {Centro Brasileiro de Pesquisas F\'isicas (CBPF), 22290-180 URCA, Rio de Janeiro (RJ), Brazil} \\
$^{14}$ {Centro de Investigaciones Energ\'eticas, Medioambientales y Tecnol\'ogicas, E-28040 Madrid, Spain} \\
$^{15}$ {Departament de F\'isica, and CERES-IEEC, Universitat Aut\`onoma de Barcelona, E-08193 Bellaterra, Spain} \\
$^{16}$ {Universit\`a di Pisa and INFN Pisa, I-56126 Pisa, Italy} \\
$^{17}$ {Department for Physics and Technology, University of Bergen, Norway} \\
$^{18}$ {INFN MAGIC Group: INFN Sezione di Torino and Universit\`a degli Studi di Torino, I-10125 Torino, Italy} \\
$^{19}$ {INFN MAGIC Group: INFN Sezione di Catania and Dipartimento di Fisica e Astronomia, University of Catania, I-95123 Catania, Italy} \\
$^{20}$ {INFN MAGIC Group: INFN Sezione di Bari and Dipartimento Interateneo di Fisica dell'Universit\`a e del Politecnico di Bari, I-70125 Bari, Italy} \\
$^{21}$ {Croatian MAGIC Group: University of Rijeka, Faculty of Physics, 51000 Rijeka, Croatia} \\
$^{22}$ {Universit\"at W\"urzburg, D-97074 W\"urzburg, Germany} \\
$^{23}$ {Technische Universit\"at Dortmund, D-44221 Dortmund, Germany} \\
$^{24}$ {University of Geneva, Chemin d'Ecogia 16, CH-1290 Versoix, Switzerland} \\
$^{25}$ {Japanese MAGIC Group: Physics Program, Graduate School of Advanced Science and Engineering, Hiroshima University, 739-8526 Hiroshima, Japan} \\
$^{26}$ {Deutsches Elektronen-Synchrotron (DESY), D-15738 Zeuthen, Germany} \\
$^{27}$ {Armenian MAGIC Group: ICRANet-Armenia, 0019 Yerevan, Armenia} \\
$^{28}$ {University of Lodz, Faculty of Physics and Applied Informatics, Department of Astrophysics, 90-236 Lodz, Poland} \\
$^{29}$ {Croatian MAGIC Group: Josip Juraj Strossmayer University of Osijek, Department of Physics, 31000 Osijek, Croatia} \\
$^{30}$ {Finnish MAGIC Group: Finnish Centre for Astronomy with ESO, Department of Physics and Astronomy, University of Turku, FI-20014 Turku, Finland} \\
$^{31}$ {Japanese MAGIC Group: Department of Physics, Tokai University, Hiratsuka, 259-1292 Kanagawa, Japan} \\
$^{32}$ {Universit\`a di Siena and INFN Pisa, I-53100 Siena, Italy} \\
$^{33}$ {Saha Institute of Nuclear Physics, A CI of Homi Bhabha National Institute, Kolkata 700064, West Bengal, India} \\
$^{34}$ {Inst. for Nucl. Research and Nucl. Energy, Bulgarian Academy of Sciences, BG-1784 Sofia, Bulgaria} \\
$^{35}$ {Japanese MAGIC Group: Department of Physics, Yamagata University, Yamagata 990-8560, Japan} \\
$^{36}$ {Finnish MAGIC Group: Space Physics and Astronomy Research Unit, University of Oulu, FI-90014 Oulu, Finland} \\
$^{37}$ {Japanese MAGIC Group: Chiba University, ICEHAP, 263-8522 Chiba, Japan} \\
$^{38}$ {Japanese MAGIC Group: Institute for Space-Earth Environmental Research and Kobayashi-Maskawa Institute for the Origin of Particles and the Universe, Nagoya University, 464-6801 Nagoya, Japan} \\
$^{39}$ {Croatian MAGIC Group: Ru\dj{}er Bo\v{s}kovi\'c Institute, 10000 Zagreb, Croatia} \\
$^{40}$ {Japanese MAGIC Group: Department of Physics, Kyoto University, 606-8502 Kyoto, Japan} \\
$^{41}$ {INFN MAGIC Group: INFN Roma Tor Vergata, I-00133 Roma, Italy} \\
$^{42}$ {Japanese MAGIC Group: Department of Physics, Konan University, Kobe, Hyogo 658-8501, Japan} \\
$^{43}$ {also at International Center for Relativistic Astrophysics (ICRA), Rome, Italy} \\
$^{44}$ {also at Port d'Informaci\'o Cient\'ifica (PIC), E-08193 Bellaterra (Barcelona), Spain} \\
$^{45}$ {also at Institute for Astro- and Particle Physics, University of Innsbruck, A-6020 Innsbruck, Austria} \\
$^{46}$ {also at Department of Physics, University of Oslo, Norway} \\
$^{47}$ {also at Dipartimento di Fisica, Universit\`a di Trieste, I-34127 Trieste, Italy} \\
$^{48}$ {Max-Planck-Institut f\"ur Physik, D-80805 M\"unchen, Germany} \\
$^{49}$ {also at INAF Padova} \\
$^{50}$ {Japanese MAGIC Group: Institute for Cosmic Ray Research (ICRR), The University of Tokyo, Kashiwa, 277-8582 Chiba, Japan} \\
$^{51}$ {also at Universitäts-Sternwarte M\"unchen, Fakult\"at f\"ur Physik, Ludwig-Maximilian-Universität M\"unchen, Scheinerstr. 1, 81679 M\"unchen, Germany} \\
$^{52}$ {INAF-Osservatorio Astronomico di Padova, Vicolo dell’Osservatorio 5, I-35122, Padova, Italy} \\


\appendix

\section{Observables in intensity interferometry}\label{sec:observables}

The central quantity in astronomical interferometry is the visibility
\begin{equation}
V(\vec{x}) \propto \int \! S_\lambda(\vec{s}) \,
e^{-\frac{2\pi i}\lambda \, \vec{b}\cdot\vec{s}} \,
d^2\vec{s}
\end{equation}
which goes back to \cite{1921ApJ....53..249M}, but is nowadays
understood as the spatial Fourier transform of the source brightness
distribution.  Here $\vec{s}$ is an angular location on the sky,
$S_\lambda(\vec{s})$ is the spectral brightness at that location,
$\vec{x}$ (called the baseline) is a vector at the observatory
transverse to the line of sight, and $V$ is normalised such that
$V(0)=1$.  If $S_\lambda(\vec{s})$ is a uniform disc of diameter
$\theta_{UD}$, $V(d)$ reduces to the well-known Airy pattern, shown in equation (\ref{eq:Bessel}).

\begin{equation}\label{eq:hbt}
c(\vec{x}) \equiv \frac{\langle I_1 I_2 \rangle} {\langle I_1 \rangle
  \langle I_2 \rangle} - 1 = \frac1{2\Delta\nu\Delta t}{|V(\vec{x})|^2}
\end{equation}
where $\Delta t$ is the time resolution and
$\Delta\nu=(c/\lambda)\Delta\lambda$ is the frequency bandwidth.  The
time resolution is sometimes expressed as the reciprocal of an
electronic bandwidth $\Delta f$.

Derivations of the relation (\ref{eq:hbt}) are given in several
sources \citep{HBbook1974,1998AcPPB..29.1839B}.  Here we draw
attention to a few essential points: (i)~intensity interferometry in
effect replaces the requirement of coherent optical paths with the
requirement of ultrafast photon counting; (ii)~an ideal intensity
interferometer would measure the energy and arrival time of every
photon as accurately as the uncertainty principle allows;
(iii)~current technology is orders of magnitude from ideal, and hence
practical intensity interferometers have very low S/N; (iv)~the low
S/N can be mitigated by collecting more photons.

Note that the intensity correlation shown in eq. (\ref{eq:hbt}) differs from the
Pearson correlation
\begin{equation}
\rho = \frac{\langle I_1 I_2 \rangle}
{\sqrt{\langle I_1^2 \rangle \langle I_2^2 \rangle}}
\end{equation}
In this work, as described in Section~\ref{sec:analysis}, $\rho$ is
computed in the hardware and includes contributions from both the
source and the night sky background. Eq.~(\ref{eq:contrast}) is needed to
remove the latter and then convert extract the desired correlation $c$
from $\rho$.


\bsp	
\label{lastpage}
\end{document}